\documentclass[
 pra,
 reprint,
 longbibliography,
 superscriptaddress,
]{revtex4-2}

\usepackage[utf8]{inputenc}
\usepackage{amsmath}
\usepackage{amssymb}
\usepackage{amsthm}
\usepackage{physics}
\usepackage{graphicx}

\usepackage{xcolor}
\usepackage{hyperref}
\usepackage{siunitx}

\begin{document}

\title{Parity-encoding-based quantum computing with Bayesian error tracking}

\author{Seok-Hyung Lee}
\affiliation{Department of Physics and Astronomy, Seoul National University, Seoul 08826, Republic of Korea}
\author{Srikrishna Omkar}
\affiliation{ORCA Computing, Toronto M6P3T1, Canada}
\author{Yong Siah Teo}
\affiliation{Department of Physics and Astronomy, Seoul National University, Seoul 08826, Republic of Korea}
\author{Hyunseok Jeong}
\affiliation{Department of Physics and Astronomy, Seoul National University, Seoul 08826, Republic of Korea}

\begin{abstract}
Measurement-based quantum computing (MBQC) in linear optical systems is promising for near-future quantum computing architecture.
However, the nondeterministic nature of entangling operations and photon losses hinder the large-scale generation of graph states and introduce logical errors.
In this work, we propose a linear optical topological MBQC protocol employing multiphoton qubits based on the parity encoding, which turns out to be highly photon-loss tolerant and resource-efficient even under the effects of nonideal entangling operations that unavoidably corrupt nearby qubits.
For the realistic error analysis, we introduce a Bayesian methodology, in conjunction with the stabilizer formalism, to track errors caused by such detrimental effects.
We additionally suggest a graph-theoretical optimization scheme for the process of constructing an arbitrary graph state, which greatly reduces its resource overhead.
Notably, we show that our protocol is advantageous over several other existing approaches in terms of fault-tolerance, resource overhead, or feasibility of basic elements.
\end{abstract}

\maketitle

\section{Introduction}

Photonic qubits are a promising candidate for quantum computing with advantages such as long decoherence time even at room temperature.
Among different encoding schemes, those of dual-rail allow one to detect photon losses by counting the total photon number and manipulate and measure single qubits \emph{via} linear optical elements and photodetectors \cite{ralph2010optical}.
A representative way to achieve universal quantum computing in linear optical systems is measurement-based quantum computing (MBQC) \cite{raussendorf2001one, raussendorf2003measurement} processed by single-qubit measurements on a multi-qubit \textit{graph state}.
In particular, a family of graph states called Raussendorf-Harrington-Goyal (RHG) lattices \cite{raussendorf2006fault, raussendorf2007topological, fowler2009topological} permits universal fault-tolerant quantum computing \cite{herr2018lattice, brown2020universal, bombin2021logical}.

The generation of RHG lattices, which is a significant challenge for realizing fault-tolerant optical MBQC, can be done by entangling multiple small resource states with fusions of types~I and/or~II \cite{browne2005resource}.
Both types of fusions are not ideal in linear optics because of theoretical limitations and environmental factors such as photon losses.
Fusion success rates cannot exceed 50\% without additional resources \cite{braunstein1995measurement} for single-photon qubits, which is far too insufficient to implement MBQC \cite{auger2018fault}.
There exist several types of approaches to overcome this shortcoming. 
Some examples include (i) different types of encoding strategies with coherent states \cite{jeong2001quantum, jeong2002efficient}, hybrid qubits \cite{omkar2020resource, omkar2021highly}, and multiphoton qubits \cite{lee2015nearly, omkar2022all} that significantly improve error thresholds and resource overheads \cite{omkar2022all}, (ii) adding ancillary photons to boost the success rate of a type-II fusion to 75\% \cite{grice2011arbitrarily, ewert2014efficient}, which enables MBQC with the renormalization method \cite{herr2018local}, (iii) redundant structures added to resource states to replace a single fusion by multiple fusion attempts \cite{fujii2010fault, li2010fault, li2015resource}, and (iv) the use of squeezing for teleportation channels \cite{takeda2013deterministic} or inline-processes \cite{zaidi2013beating, kilmer2019boosting}.

Previous studies frequently treated fusion failures with bond disconnection \cite{gimeno2015from, zaidi2015near, pant2019percolation} or qubit removals \cite{auger2018fault, herr2018local, omkar2020resource, omkar2022all}.
However, to accurately evaluate the performance of computing protocols, the detrimental effects of nonideal fusions affecting nearby qubits should be analyzed more rigorously.
In this work, we study how nonideal fusions corrupt stabilizers and how errors arising from such corruption can be tracked during the generation of graph states.
Using a Bayesian approach and the stabilizer formalism, we can now assign error rates with strong posterior evidence from measurement data on certain qubits in the final lattice, thereby enabling much more realistic error simulations and adaptive decoding of syndromes.

We then propose a linear-optical fault-tolerant MBQC protocol termed a \textit{parity-encoding-based topological quantum computing} (PTQC), which employs the parity encoding \cite{ralph2005loss} and concatenated Bell-state measurement (CBSM) \cite{lee2019fundamental}.
The protocol requires on-off or single-photon resolving detectors, optical switches, delay lines, and three-photon Greenberger-Horne-Zeilinger (GHZ-3) states that can be generated with linear optics.
(A single-photon resolving detector discriminates between zero, one, and more than one photons entering the detector.)
We analyze the loss-tolerance of the protocol while exhaustively tracking the detrimental effects of nonideal fusions.
The resource overhead in terms of the number of required GHZ-3 states is also investigated.
To minimize it, we introduce a graph-theoretical method for optimizing the process of constructing resource states, which is generalizable for other MBQC schemes.
By comparing PTQC with three other known approaches using single-photon qubits with fusions assisted by ancillary photons, using simple repetition codes, and using redundant tree graphs, we show that our protocol is advantageous over these protocols in terms of fault-tolerance, resource overheads, or feasibility of basic elements.

\begin{figure}[t!]
	\centering
	\includegraphics[width=\columnwidth]{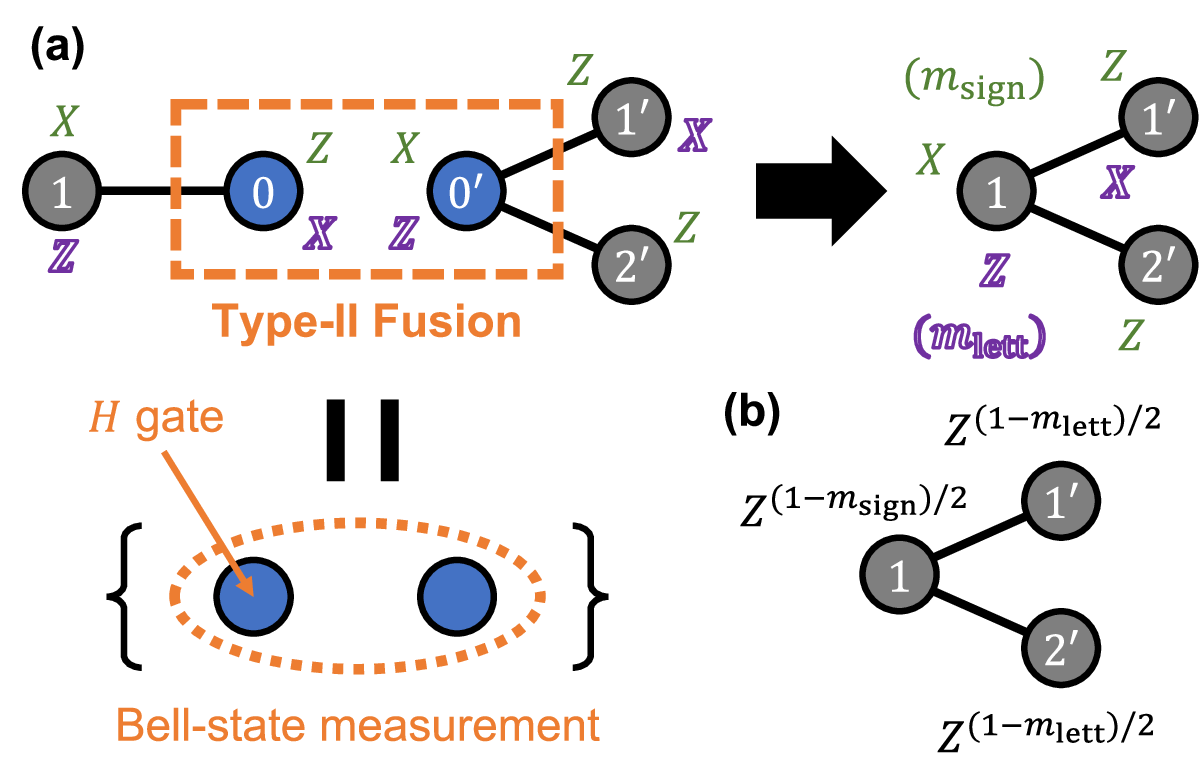}
	\caption{
        \textbf{Example of a type-II fusion.}
	    A type-II fusion is done by measuring $Z_0 X_{0'}$ and $X_0 Z_{0'}$ on the two graph states.
	    In (a), two stabilizers (green and purple operators) become those of the resulting graph state up to sign factors (the sign or letter outcome $m_\mathrm{sign}$, $m_\mathrm{lett}$ of the BSM) after the fusion.
	    The final state is the graph state shown in (b), where the presented Pauli-$Z$ operators are applied.
	}
	\label{fig:fusion}
\end{figure}

\subsection{Preliminaries}
We denote the four Bell states by $\ket{\phi^\pm} := \ket{0}\ket{0} \pm \ket{1}\ket{1}$ and $\ket{\psi^\pm} := \ket{0}\ket{1} \pm \ket{1}\ket{0}$ (normalization coefficients are omitted) and call ``$\pm$'' its \textit{sign} and ``$\phi$'' or ``$\psi$'' its \textit{letter}.
An ideal Bell-state measurement (BSM) entails the measurements of $X \otimes X$ and $Z \otimes Z$ on two qubits, whose outcomes are addressed as its \textit{sign and letter outcomes}, respectively.
We use the polarization of photons as the degree of freedom to encode quantum information and denote the horizontally (vertically) polarized single-photon state by $\ket{\textsc{h}}$ ($\ket{\textsc{v}}$).

For a given graph $G$ of qubits, a graph state $\ket{G}$ is defined as the state stabilized by $S_v := X_v \prod_{v' \in N(v)} Z_{v'}$ (that is, $S_v \ket{G} = \ket{G}$) for each vertex $v$, where $X_v$ and $Z_v$ are respectively Pauli-$X$ and $Z$ operators on the qubit $v$ and $N(v)$ is the set of the vertices connected with $v$.
$\ket{G}$ can be generated by placing a qubit initialized as $\ket{+} := \ket{\textsc{h}} + \ket{\textsc{v}}$ on each vertex of $G$ and applying a controlled-$Z$ gate on every pair of qubits connected by an edge in $G$.
However, since the direct implementation of a controlled-$Z$ gate for photonic qubits demands multi-photon interaction, linear optical MBQC typically takes an approach to construct a graph state by merging multiple small resource graph states via fusion operations \cite{browne2005resource, kieling2007percolation, fujii2010fault, li2010fault, gimeno2015from, li2015resource, zaidi2015near, herr2018local, pant2019percolation, omkar2020resource, omkar2022all}.

Among the two types of fusions \cite{browne2005resource}, we only consider type II since type I may convert photon losses into unheralded errors \cite{li2015resource}.
A type-II fusion is done by measuring $X \otimes Z$ and $Z \otimes X$ on two qubits.
In practice, it is realized by applying the Hadamard gate on one of the qubits and then performing a BSM on them.
For two qubits $(v_1, v_2)$, if $\qty{v_1} \cup N\qty(v_1)$ and $\qty{v_2} \cup N\qty(v_2)$ are disjoint, the effect of a fusion on the qubits is to connect (disconnect) every possible pair of disconnected (connected) qubits, one from $N\qty(v_1)$ and the other from $N\qty(v_2)$, up to several Pauli-$Z$ operators determined by the BSM outcome.
These Pauli-$Z$ operators are compensated by updating the Pauli frame \cite{knill2005quantum} classically.
This effect can be checked by tracking stabilizers, as shown in the example of Fig.~\ref{fig:fusion}(a).
Here, the stabilizer $X_1 Z_0 X_{0'} Z_{1'} Z_{2'}$ (colored in green) before the fusion is transformed into $m_\mathrm{sign} X_1 Z_{1'} Z_{2'}$ after the fusion, where $m_\mathrm{sign} \in \qty{\pm 1}$ is the sign outcome of the BSM if the Hadamard gate is applied on qubit $0$. 
The other two stabilizers $Z_1 X_0 Z_{0'} X_{1'}$ (colored in purple) and $Z_1 X_0 Z_{0'} X_{2'}$ that commute with the fusion can be transformed in similar ways.
Consequently, the marginal state on the unmeasured qubits is equal to the merged graph state up to several Pauli-$Z$ operators, as presented in Fig.~\ref{fig:fusion}(b).

We consider errors of qubits in the ``vacuum'' measured in the $X$-basis, which occupies most of the area in the RHG lattice \cite{raussendorf2006fault}; thus, $X$-errors do not affect the results.
Henceforth, every error mentioned is a $Z$-error.

\section{Results}
\subsection{Bayesian error tracking for nonideal fusions}

\begin{figure}[t!]
	\centering
	\includegraphics[width=\columnwidth]{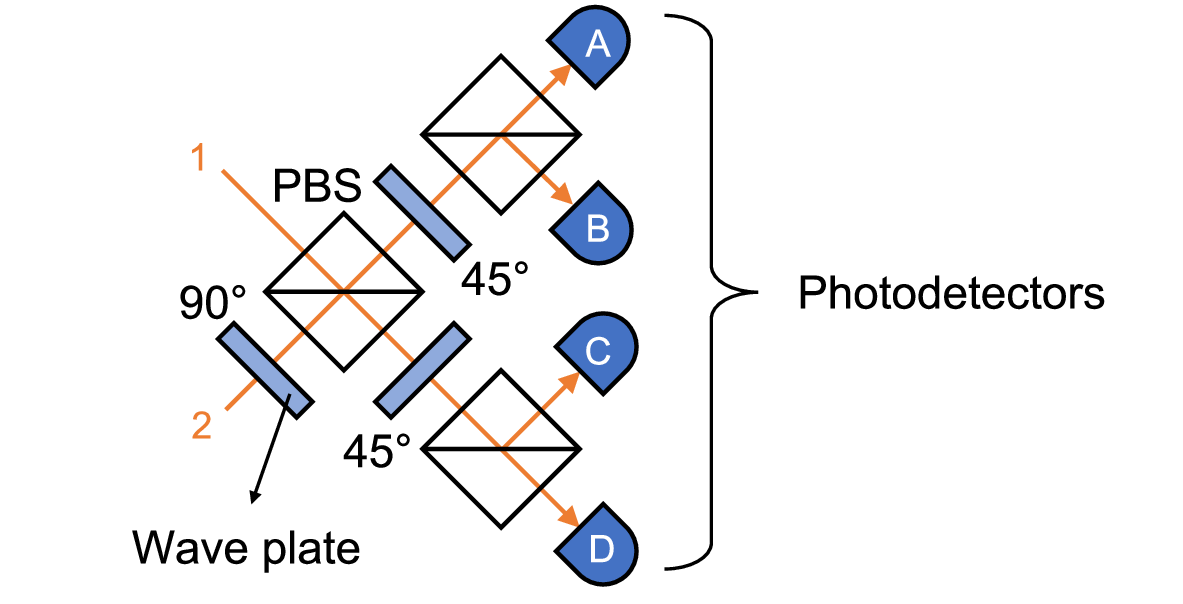}
	\caption{
	\textbf{BSM scheme for single-photon polarization qubits.}
    It uses three polarizing Beam splitters (PBSs), \ang{90} and \ang{45} wave plates, and four (A--D) photodetectors (single-photon resolving or on-off detectors).
	A PBS transmits (reflects) photons polarized horizontally (vertically).
	The scheme distinguishes $\ket{\psi^\pm}$: $\ket{\psi^+}$ if detectors $(\mathrm{A}, \mathrm{C})$ or $(\mathrm{B}, \mathrm{D})$ detect one photon respectively and $\ket{\psi^-}$ if detectors $(\mathrm{A}, \mathrm{D})$ or $(\mathrm{B}, \mathrm{C})$ detect one photon respectively.
	If otherwise, it fails or detects a loss, which can be distinguished by the total number of detected photons if single-photon resolving detectors are used.
	Two distinguishable Bell states can be chosen by putting or removing wave plates appropriately before the first PBS.
	}
	\label{fig:physical_bsm_scheme}
\end{figure}

We now introduce the methodology to track the errors caused by nonideal fusions.
Let us revisit the example in Fig.~\ref{fig:fusion}, supposing that the qubits are single-photon polarization ones and there are no photon losses.
Then a BSM can discriminate between only two Bell states (say, $\ket{\psi^\pm}$) among the four without additional resources \cite{lutkenhaus1999bell}; see Fig.~\ref{fig:physical_bsm_scheme} for the scheme.
The intact final state $\ket{C_\mathrm{f}}$ is obtained only when the BSM succeeds.
When the BSM fails (which is heralded), $m_\mathrm{lett}$ is determined while $m_\mathrm{sign}$ is left completely ambiguous.
In other words, the posterior probability that the input state is $\ket{\phi^\pm}$ for the obtained photodetector outcomes is equal for both signs ($\pm$), assuming that the four Bell states have the same prior probability.
This assumption can be justified by the fact that the marginal state on qubits $0$ and $0'$ before the fusion is maximally mixed; see Supplementary Note~1 for the proof.
Therefore, we fix the value of $m_\mathrm{lett}$ while randomly assign that of $m_\mathrm{sign}$.
Then, the operator $m_\mathrm{sign} X_1 Z_{1'} Z_{2'}$, which is originally a stabilizer of $\ket{C_\mathrm{f}}$, gives $\pm 1$ randomly when it is measured after the failed BSM.
Whereas, the other two stabilizers $m_\mathrm{lett} Z_1 X_{1'}$ and $m_\mathrm{lett} Z_1 X_{2'}$ are left undamaged.
The key point is that this situation is equivalent to a 50\% chance of an erroneous qubit~1 in $\ket{C_\mathrm{f}}$ in terms of stabilizer statistics.
In other words, both situations give the same statistics if the stabilizers of $\ket{C_\mathrm{f}}$ are measured; thus, every process in MBQC described with the stabilizer formalism works in the same way.

Generally, a nonideal BSM gives one of the possible outcomes and the posterior probability of each Bell state for the outcome can be calculated with the Bayesian theorem, assuming the equal prior probabilities of the Bell states.
Accordingly, the Bell state with the highest posterior probability is selected as the result of the BSM, and the probability $q_\mathrm{sign}$ ($q_\mathrm{lett}$) that the selected sign (letter) is wrong can be obtained as well.
These error probabilities are ``propagated'' into nearby qubits in a way that the stabilizer statistics are preserved.
For example, if the fusion in Fig.~\ref{fig:fusion} is nonideal in such a way, it is equivalent to qubit~1 having an error with probability $q_\mathrm{sign}$ and qubits $1'$ and $2'$ having correlated errors with probability $q_\mathrm{lett}$.
We term a qubit with a nonzero error rate \textit{deficient}.

Additionally, if a qubit participating in a fusion is erroneous, this error is propagated to the qubits on the opposite side.
For example, an erroneous qubit~0 in Fig.~\ref{fig:fusion} induces an error in the $X_0 Z_{0'}$ measurement, which is equivalent to erroneous qubits $1'$ and $2'$.

The above error tracking methodology can be utilized for accurate and effective error simulations.
The method can precisely locate qubits affected by unsuccessful fusions, which is closer to reality than simple bond disconnection or qubit removal.
Since unsuccessful fusions are now regarded as Pauli error sources, we no longer need lattice deformation and the construction of supercheck operators \cite{barrett2010fault, auger2018fault}.
Instead, the error probabilities on individual qubits are employed for decoding syndromes in an adaptive manner (with decoders such as the \textit{weighted} minimum-weight perfect matching one), which may be particularly effective if the probabilities are between 0 and $1/2$ since regarding such errors as just removal of qubits is a loss of information.

\subsection{Building an RHG lattice}

\begin{figure}[t!]
	\centering
	\includegraphics[width=\columnwidth]{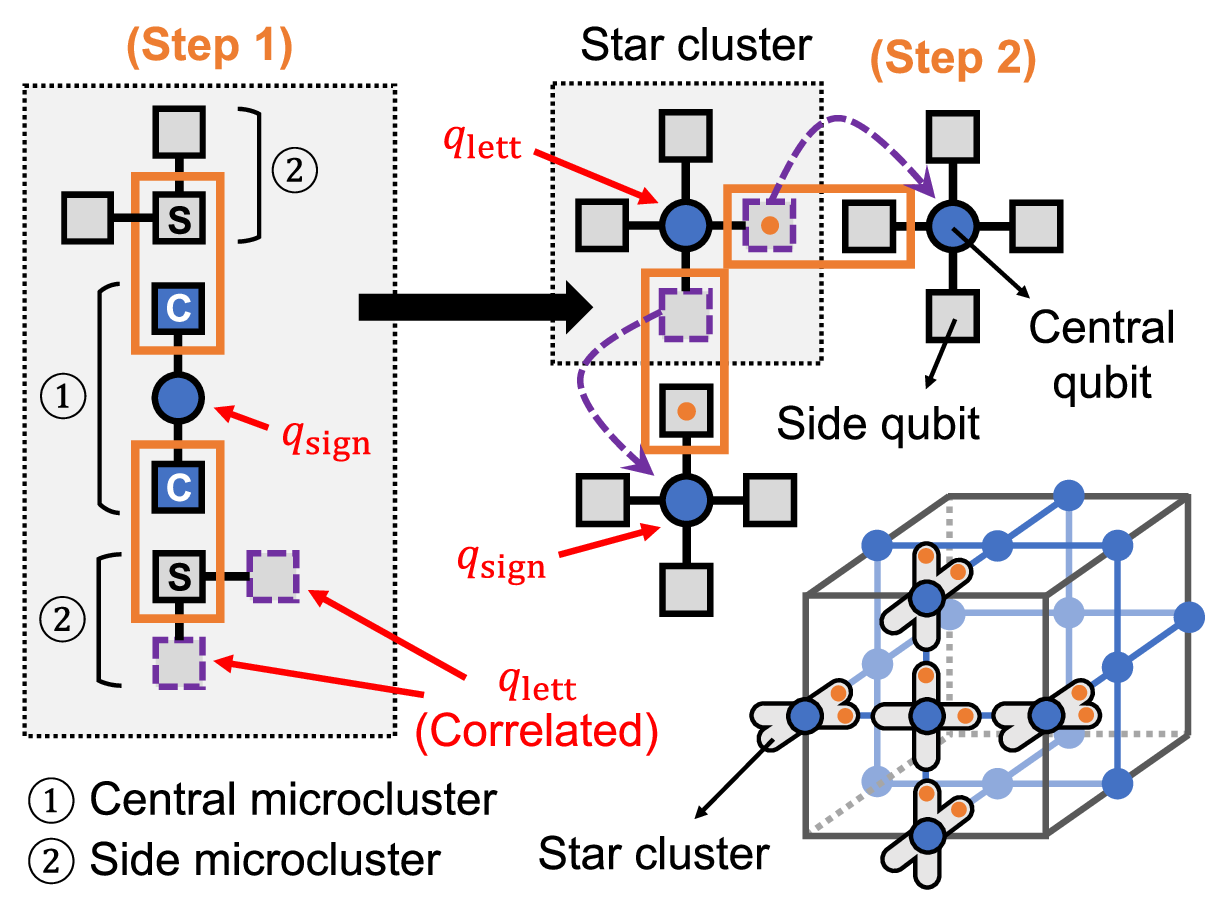}
	\caption{
        \textbf{Lattice building process with microclusters.}
        The orange boxes indicate fusions.
        In step 1, side and central microclusters are fused to form a star cluster.
        The locations of the Hadamard gates are marked as ``C'' (``S'') for the HIC (HIS) configuration.
        In step 2, multiple star clusters are fused to form an RHG lattice.
        The macroscopic picture of step 2 in a unit cell of the lattice is depicted in the lower right.
        The locations of the Hadamard gates are marked as orange dots.
        The error probabilities of qubits assigned by one fusion in each step for the HIC configuration are written in red, where $q_\mathrm{sign}$ ($q_\mathrm{lett}$) is the sign (letter) error probability of the BSM.
        Errors in the side qubits remaining after step 1 (purple dashed squares) are propagated to central qubits during step 2 (purple dashed arrows).
	}
	\label{fig:building_lattice}
\end{figure}

An RHG lattice can be built with two types of linear three-qubit graph states called \textit{central and side microclusters} \cite{gimeno2015from, herr2018local}.
The process is composed of two steps (see Fig. \ref{fig:building_lattice}):
In step~1, a central microcluster and two side microclusters are merged by two fusions to form a five-qubit graph state named a \textit{star cluster} composed of one \textit{central qubit} and four \textit{side qubits}.
In step~2, the side qubits of star clusters are fused to form an RHG lattice.
Eventually, the lattice includes only the central qubits, which are measured in appropriate bases for MBQC.
For step~2, we consider two options: (i) Star clusters with successful step-1 fusions may be post-selected, or (ii) all generated star clusters are used regardless of the fusion results.
The locations of the Hadamard gates during fusions (called \textit{$H$-configuration}) may be chosen arbitrarily.
Here, we define two specific $H$-configurations: \textit{Hadamard-in-center (HIC)} and \textit{Hadamard-in-side (HIS)}.
In the HIC (HIS) configuration, the Hadamard gates in step~1 are applied on qubits in the central (side) microclusters, as shown in Fig. \ref{fig:building_lattice}.
Whereas the Hadamard gates in step~2 are arranged in the same pattern for both configurations.

Nonideal fusions during lattice building render some central qubits in the final lattice deficient, as shown in Fig.~\ref{fig:building_lattice} when the HIC configuration is used.
When the HIS configuration is used, the positions of $q_\mathrm{sign}$ and $q_\mathrm{lett}$ in the figure are swapped.
Note that errors in the side qubits are propagated to the nearest central qubits after step~2.
Correlation between the sign and letter errors of a fusion, if any, can be neglected if the primal and dual lattices are considered separately, since these errors respectively affect primal and dual \cite{raussendorf2006fault} qubits (or vice versa).

\subsection{Noise model}
For analyzing the following linear optical quantum computing protocols, we consider a noise model where each photon suffers an independent loss with probability $\eta$, which arises from imperfections throughout the protocol: GHZ-3 states (which are initial resource states), delay lines, optical switches, and photodetectors.
We assume that noise that cannot be modeled with photon losses such as dark counts is negligible.
Note that not only nonideal fusions but also photon losses in central qubits, which are detectable by on-off detectors, may incur deficiency.
If the measurement outcome of a central qubit cannot be determined due to photon losses, we select the outcome randomly and assign an error rate of 50\% to the qubit.

\subsection{Parity-encoding-based topological quantum computing}
We introduce the new linear-optical parity-encoding-based topological quantum computing (PTQC) protocol, where fusion success rates are boosted by using multiphoton qubits for all qubits that participate in fusions and single-photon polarization encoding is used for central qubits.
The parity encoding \cite{ralph2005loss} is employed for the multiphoton qubits, which are fused by CBSM \cite{lee2019fundamental}.
On-off or single-photon resolving detectors are used as photodetectors, and GHZ-3 states, which can be generated linear-optically \cite{varnava2008how}, are regarded as basic resource states.
The $(n, m)$ parity encoding defines a basis as 
\begin{align}
    \ket{0_L} := \ket{+^{(m)}}^{\otimes n}, \qquad \ket{1_L} := \ket{-^{(m)}}^{\otimes n},
    \label{eq:lattice_level_def}
\end{align}
where
\begin{align}
    \ket{\pm^{(m)}} := \qty(\ket{\textsc{h}} + \ket{\textsc{v}})^{\otimes m} \pm \qty(\ket{\textsc{h}} - \ket{\textsc{v}})^{\otimes m}.
    \label{eq:block_level_def}
\end{align}
The Hilbert space has a hierarchical structure composed of three levels: the lattice, block, and physical levels with respective bases $\qty{\ket{0_L}, \ket{1_L}}$, $\qty{\ket{\pm^{(m)}}}$, and $\qty{\ket{\textsc{h}}, \ket{\textsc{v}}}$.
In the original CBSM scheme \cite{lee2019fundamental}, a BSM of a certain level is decomposed into multiple BSMs of one level below.
Our current CBSM scheme slightly differs from the original one in the following two areas: 
{(i) We consider two types of photodetectors: single-photon resolving and on-off detectors.
A physical-level BSM can discriminate between a photon loss and failure only if single-photon resolving detectors are used.}
(ii) The letter outcome of a lattice-level BSM is obtained by a weighted majority vote of block-level letter outcomes.
See the Methods section for details of the CBSM scheme and its error rates.

\begin{figure}[t!]
	\centering
	\includegraphics[width=\columnwidth]{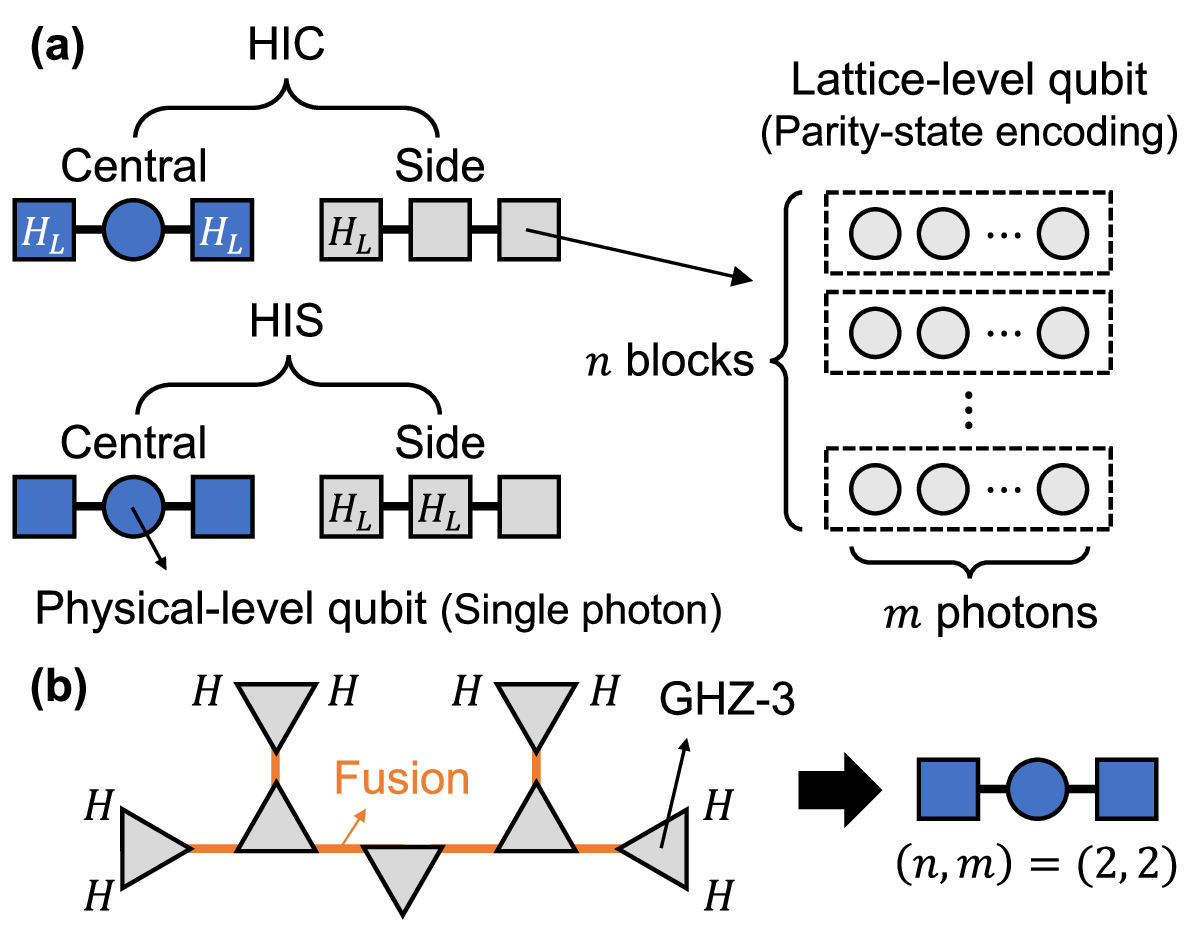}
	\caption{
        \textbf{Structure and generation of post-$H$ microclusters for PTQC.}
	    (a) Schematic of central and side post-$H$ microclusters used in PTQC for the two $H$-configurations, HIC and HIS.
	    The marks ``$H_L$'' indicate the locations of the lattice-level Hadamard gates.
	    (b) Example of a process generating a post-$H$ microcluster from GHZ-3 states.
	    Each GHZ-3 state is represented by a triangle whose vertices indicate its three photons.
	    An orange line connecting two vertices and a mark ``$H$'' next to a vertex respectively mean a fusion and Hadamard gate performed on the photon(s).
        The graph of the triangles connected with the orange lines is called a merging graph.
	}
	\label{fig:encoded_microclusters}
\end{figure}

For practical reasons, we consider generating \textit{post-$H$} microclusters (that is, the states obtained by applying several lattice-level Hadamard gates on microclusters) directly from GHZ-3 states, instead of generating microclusters first and then applying the lattice-level Hadamard gates for the fusions.
Figure~\ref{fig:encoded_microclusters}(a) depicts the central and side post-$H$ microclusters for the HIC and HIS configurations.
A post-$H$ microcluster can be generated up to several physical-level Hadamard gates by performing physical-level BSMs or fusions (referred to as \textit{merging operations}) between multiple GHZ-3 states according to a predetermined \textit{merging graph}, as shown in the example of Fig.~\ref{fig:encoded_microclusters}(b).
Note that the merging graph may be not unique for a post-$H$ microcluster.
However, each merging operation has a low success rate of less than or equal to 50\%, which may lead to extensive usage of GHZ-3 states for generating a post-$H$ microcluster successfully.
Thus, the generation process, which is determined by the merging graph and the order of the merging operations, should be adjusted carefully to minimize the resource overhead.
To optimize the merging order, our protocol utilizes a graph edge coloring algorithm, based on the idea that merging operations for non-adjacent edges can be performed simultaneously.
See the Methods section for details of the structures of post-$H$ microclusters, their generation, and the resource optimization problem.

For error simulations, we consider the logical identity gate with the length $T$ of $4d + 1$ unit cells along the simulated time axis, where $d$ is the code distance.
All the fusion outcomes are sampled from appropriate probability distributions, and the corresponding error rates are assigned to individual central qubits according to the process described earlier.
These error rates are exploited when decoding syndromes by the \textit{weighted} minimum-weight perfect matching in the PyMatching package \cite{higgott2021pymatching}.
The loss thresholds are calculated by finding the intersections of logical error rates for $d=9$ and $d=11$.
The resource overhead of PTQC is quantified by the average total number $\mathcal{N}_{p_L}$ of GHZ-3 states to achieve a target logical error rate of $p_L$ for the logical identity gate of $T = d-1$, which depends on the photon loss rate $\eta$.
See Supplementary Notes~2 and 3 for the detailed methods of error simulations and resource calculations, respectively.

\begin{figure*}[t!]
	\centering
	\includegraphics[width=\textwidth]{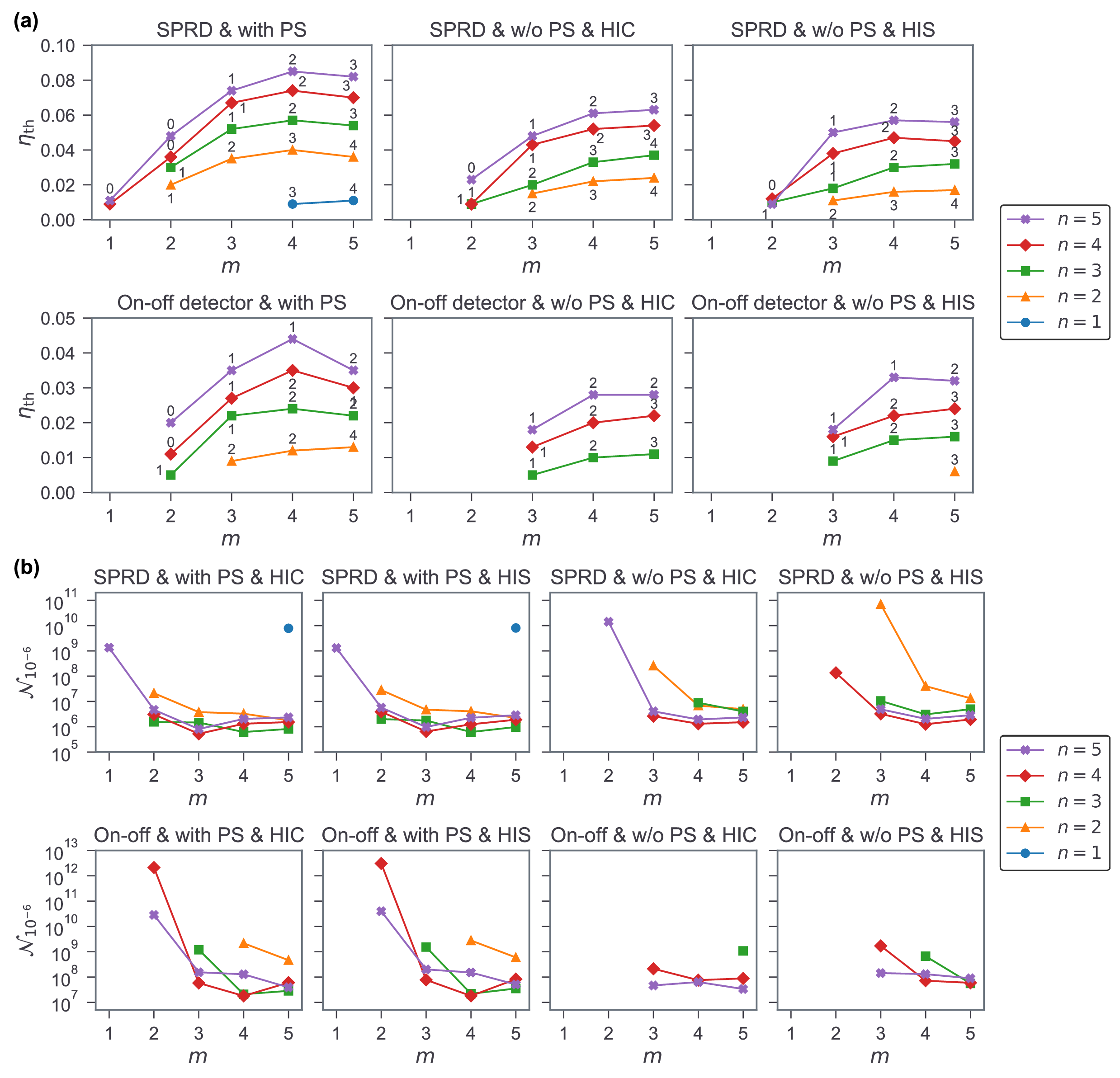}
	\caption{
        \textbf{Simulation results for PTQC.}
        (a) Loss threshold $\eta_\mathrm{th}$ and (b) resource overhead $\mathcal{N}_{10^{-6}}$ are calculated for various parameters on the encoding size $(n, m)$, the type of detectors, the post-selection (PS) of star clusters, and the $H$-configuration.
    	``SPRD'' stands for single-photon resolving detector.
        $\mathcal{N}_{10^{-6}}$ is calculated at $\eta = 0.01$.
	    The values of $j$ are chosen to maximize $\eta_\mathrm{th}$ and shown next to the data points in (a).
	    The $H$-configuration does not affect $\eta_\mathrm{th}$ when star clusters are post-selected.
    }
	\label{fig:simulation_encoding}
\end{figure*}

\begin{figure*}[t!]
	\centering
	\includegraphics[width=\textwidth]{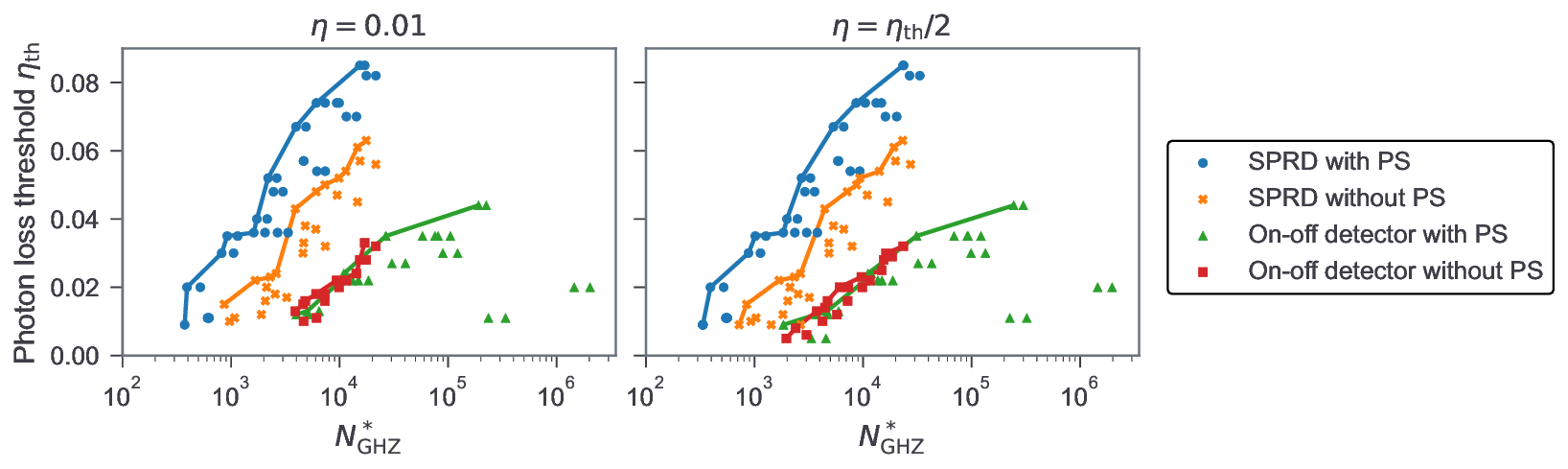}
	\caption{
	    \textbf{Photon loss thresholds $\eta_\mathrm{th}$ as a function of the number $N_\mathrm{GHZ}^*$ of GHZ-3 states required per central qubit.}
        $N_\mathrm{GHZ}^*$ is calculated at $\eta = 0.01$ or $\eta = \eta_\mathrm{th}/2$.
       ``SPRD'' stands for a single-photon resolving detector.
	    The data points correspond to different parameter settings on the type of detectors, the post-selection (PS) of star clusters, the encoding size, and the $H$-configuration, which are grouped by the first two factors.
	    The upper envelope for each of the groups is presented as a line.
	    The values of $j$ are chosen to maximize $\eta_\mathrm{th}$.
	}
	\label{fig:thres_vs_overheads}
\end{figure*}

The simulation results of the loss thresholds and the resource overheads (quantified by $\mathcal{N}_{10^{-6}}$) are respectively presented in Figs.~\ref{fig:simulation_encoding}(a) and (b) for the two types of photodetectors, the two options for the post-selection of star clusters, and the two $H$-configurations.
Figure~\ref{fig:simulation_encoding}(a) shows that, if single-photon resolving detectors are used, $\eta_\mathrm{th}$ reaches up to 8.5\% ($n=5,~m=4,~j=2$) when star clusters are post-selected and up to 6.3\% ($n=m=5,~j=3$, HIC) when they are not.
If on-off detectors are used, $\eta_\mathrm{th}$ reaches up to 4.4\% ($n=5,~m=4,~j=1$) when star clusters are post-selected and up to 3.3\% ($n=5,~m=4,~j=1$, HIS) when they are not.
The post-selection of star clusters increases the photon loss thresholds by about $1\textendash2\%\mathrm{p}$.
From Fig.~\ref{fig:simulation_encoding}(b), it is observed that the protocol using single-photon resolving detectors is most resource-efficient with $\mathcal{N}_{10^{-6}} \approx 5 \times 10^5$ ($n=4,~m=3,~j=1$, HIC) when star clusters are post-selected and with $\mathcal{N}_{10^{-6}} \approx 1 \times 10^6$ ($n=m=4,~j=2$, HIS) when they are not.
If on-off detectors are used, the protocol is most resource-efficient with $\mathcal{N}_{10^{-6}} \approx 2 \times 10^7$ ($n=m=4,~j=2$, HIC) when star clusters are post-selected and with $\mathcal{N}_{10^{-6}} \approx 3 \times 10^7$ ($n=m=5,~j=2$, HIC) when they are not.
It is worth noting that, compared to the protocol without the post-selection, the protocol with it requires fewer GHZ-3 states to achieve a target logical error rate.
In other words, further fault-tolerance obtained by using only successfully-generated star clusters leads to a positive overall effect that surpasses the negative effect caused by the increase in the number of required GHZ-3 states for one central qubit in the final lattice.

Additionally, Fig.~\ref{fig:thres_vs_overheads} presents the photon loss thresholds as a function of $N_\mathrm{GHZ}^*$ when $\eta$ is fixed to $0.01$ or variable as $\eta = \eta_\mathrm{th}/2$, which is used to calculate $\mathcal{N}_{10^{-6}}$.
aIt shows that at least about 400 GHZ-3 states are required per central qubit for PTQC to work. 
The explicit information of the data points along the upper envelope lines in the figure is presented in Supplementary Table~1.

\subsection{Comparison with other approaches}

We now compare the PTQC protocol with three other known approaches for linear optical quantum computing: (i) using single photons for all qubits with fusions assisted by ancillary photons, (ii) using simple repetition codes, and (iii) attaching redundant tree structures to replace a single fusion by multiple fusion attempts.
We show evidence that PTQC is more efficient than these approaches.

We first consider the approach of (i) that uses single photons for all qubits with fusions assisted by ancillary photons \cite{ewert2014efficient}, which has been widely studied in the context of ballistic quantum computing \cite{gimeno2015from, zaidi2015near, herr2018local, pant2019percolation}.
In these works, non-RHG lattices are considered except for Ref.~\cite{herr2018local}; however, RHG lattices should be used to enable a solid error correction, as also mentioned in Refs.~\cite{gimeno2015from, pant2019percolation}.
Moreover, in these works, the detrimental effects of failed fusions corrupting nearby qubits are not treated comprehensively; instead, they (except Ref.~\cite{herr2018local}) regard a fusion failure as removing the corresponding edge and mainly focus on finding percolation thresholds.

By analyzing the approach of (i) with our methodology for handling nonideal fusions, we can show that $p_\mathrm{f}$ should be less than about 10\% (1\%) even if $\eta$ is only 1\% when star clusters are (are not) post-selected; see Supplementary Note~4 for the detailed methods and results.
The failure rate of $10\%$ can be achieved by using the BSM scheme of $N=3$ in Ref.~\cite{ewert2014efficient} where $p_\mathrm{f} = 6.25\%$.
It requires multiple photon-number resolving detectors (PNRDs) resolving up to 16 photons and specific four- and eight-photon ancillary states that are conjectured to be impossible to obtain from single photons with linear optics \cite{ewert2014efficient}.
Moreover, our simulation does not consider the imperfectness of ancillary states and additional PNRDs; if they are considered, the requirements will be even stricter.
Therefore, it may be highly demanding to implement this MBQC protocol with linear optics.
We note that there is a possibility that the lattice renormalization method in Ref.~\cite{herr2018local} makes the protocol less demanding, which is worth investigating in future works.
However, the method has a shortcoming that the renormalized lattice may be significantly smaller than the original lattice; namely, about $20^3$ photons are consumed to generate one node \cite{herr2018local}.

Next, we consider the approach of (ii) that uses simple repetition codes, which is covered in our previous work \cite{omkar2022all}.
In Ref.~\cite{omkar2022all}, the photon loss thresholds and resource overheads are analyzed in detail, but a rigorous analysis of the effects of nonideal fusions like that done for PTQC is lacking.
It is observed that the recalculated photon loss thresholds are lower than the values reported in \cite{omkar2022all}, which shows that PTQC is much more loss-tolerant than this approach.
For example, if each central qubit consists of two photons, the obtained optimal photon loss threshold is 0.97\% (0.40\%) when star clusters are (are not) post-selected, while the reported value is 3.3\% (3.1\%).
Moreover, encoding central qubits with the repetition code does not improve the performance significantly.
See Supplementary Note~5 for details.

Lastly, we compare PTQC with the approach of (iii) that utilizes redundant tree structures on graph states.
Such an approach also has been actively investigated \cite{fujii2010fault, li2010fault, li2015resource}, among which Ref.~\cite{li2015resource} presents the current most advanced version of the protocol where an RHG lattice is constructed by entangling multiple GHZ-3 states like PTQC.
There, at least $\sim 2 \times 10^5$ photodetectors are required per data qubit to achieve a positive photon loss threshold with single-photon resolving detectors, while PTQC requires at least $\sim 7 \times 10^4$ photodetectors per data qubit (see the Methods section for the calculation).
Hence, PTQC shows about a twofold improvement in resource efficiency compared to the protocol in Ref.~\cite{li2015resource}.
Furthermore, we have shown that PTQC also operates with on-off detectors, while the protocol in Ref.~\cite{li2015resource} is currently unclear whether it is possible.
Nevertheless, further work will be required to compare their performance (especially their fault-tolerance) rigorously and comprehensively.

\section{Discussion}
In this work we address the problem of overcoming the negative effects of nonideal fusions and photon losses during linear-optical measurement-based quantum computing (MBQC).
We first introduced a Bayesian methodology for tracking errors caused by nonideal fusions during the construction of graph states, which enables accurate and effective error simulations.
We then proposed the \textit{parity-encoding-based topological quantum computing} (PTQC) protocol that uses the parity encoding and concatenated Bell-state measurement, which turns out to have a high loss threshold of at most $\sim8.5\%$.
Moreover, logical error rates near $10^{-6}$ can be achieved using about $10^6$ or fewer three-photon Greenberger-Horne-Zeilinger states (GHZ-3) states in total when the photon loss rate is 1\%, which outperforms other known linear optical computing protocols \cite{omkar2022all}.
We presented comprehensive and systematic methods to construct a graph state from GHZ-3 states, including the graph-theoretical algorithm that can minimize the resource overhead efficiently.

Additionally, we investigated three other known approaches that respectively use single-photon qubits with fusions assisted by ancillary photons, simple repetition codes, and redundant tree graphs.
We verified that the first two are highly demanding compared to PTQC due to low photon loss thresholds or hard-to-implement requirements such as photodetectors that can resolve many photons.
Compared to the third approach, we showed that PTQC has a twofold improvement in terms of the resource overhead required for the loss threshold to be positive, although additional work will be necessary to compare their fault-tolerance as well.

One may apply the Bayesian error tracking method to other encoding schemes or decoding algorithms (such as the union-find decoder \cite{delfosse2021almost}) to improve fault-tolerance or resource overheads.
More careful consideration of component-wise errors, including both heralded photon losses and unheralded errors (such as dark counts on photodetectors), shall give rise to more realistic analyses.
Resource analysis will be more comprehensive if other factors such as the number of optical switches or the lengths of delay lines are considered.
In particular, one trial of CBSM may require optical switches to change the types ($B_\psi$, $B_+$, and $B_-$) of the physical-level BSMs.
Our graph-theoretical optimization scheme for generating graph states can be applied to arbitrary graph states as well as microclusters for PTQC.
It will be interesting future work to investigate the resource reduction effect of this scheme for various MBQC protocols or other applications of graph states such as quantum repeaters.
Lastly, our methods may be generalized to fusion-based quantum computing \cite{bartolucci2021fusion} that is attracting attention recently, or other MBQC protocols such as the color-code-based one \cite{lee2022universal}.

\section{Methods}

In this section, we describe the details of the PTQC protocol including the CBSM scheme, the closed-form expressions of error probabilities, the method to generate post-$H$ microclusters, and the resource optimization problem.

\subsection{Bell states for the parity encoding}

For the lattice, block, and physical levels of the $(n,m)$ parity encoding, the Bell states are respectively defined as 
\begin{align*}
    &\begin{cases}
        \ket{\Phi^\pm} := \ket{0_L} \ket{0_L} \pm \ket{1_L} \ket{1_L}, \\
        \ket{\Psi^\pm} := \ket{0_L} \ket{1_L} \pm \ket{1_L} \ket{0_L},
    \end{cases} \\
    &\begin{cases}
        \ket{\phi_{(m)}^\pm} := \ket{+^{(m)}} \ket{+^{(m)}} \pm \ket{-^{(m)}} \ket{-^{(m)}}, \\
        \ket{\psi_{(m)}^\pm} := \ket{+^{(m)}} \ket{-^{(m)}} \pm \ket{-^{(m)}} \ket{+^{(m)}},
    \end{cases} \\
    &\begin{cases}
        \ket{\phi^\pm} := \ket{\textsc{h}} \ket{\textsc{h}} \pm \ket{\textsc{v}} \ket{\textsc{v}}, \\
        \ket{\psi^\pm} := \ket{\textsc{h}} \ket{\textsc{v}} \pm \ket{\textsc{v}} \ket{\textsc{v}},
    \end{cases}
\end{align*}
where $\ket{0_L}$, $\ket{1_L}$, and $\ket{\pm^{(m)}}$ are defined in Eqs.~\eqref{eq:lattice_level_def} and \eqref{eq:block_level_def}.
\begin{subequations}
The Bell states of each level can be decomposed into those of one level below as follows:
\begin{align}
    \ket{\Phi^\pm} &= 2^{-\frac{n-1}{2}} \sum_{l:\mathrm{even(odd)} \leq n} \mathcal{P} \qty[ \ket{\phi_{(m)}^-}^{\otimes l} \ket{\phi_{(m)}^+}^{\otimes n-l} ], \label{eq:logical_phi_decomposition} \\
    \ket{\Psi^\pm} &= 2^{-\frac{n-1}{2}}\sum_{l:\mathrm{even(odd)} \leq n} \mathcal{P} \qty[ \ket{\psi_{(m)}^-}^{\otimes l} \ket{\psi_{(m)}^+}^{\otimes n-l} ], \label{eq:logical_psi_decomposition} \\
    \ket{\phi_{(m)}^\pm} &= 2^{-\frac{m-1}{2}} \sum_{k:\mathrm{even} \leq m} \mathcal{P} \qty[ \ket{\psi^\pm}^{\otimes k} \ket{\phi^\pm}^{\otimes m-k} ], \label{eq:block_phi_decomposition} \\
    \ket{\psi_{(m)}^\pm} &= 2^{-\frac{m-1}{2}} \sum_{k:\mathrm{odd} \leq m} \mathcal{P} \qty[ \ket{\psi^\pm}^{\otimes k} \ket{\phi^\pm}^{\otimes m-k} ], \label{eq:block_psi_decomposition}
\end{align}
where $\mathcal{P}[\cdot]$ means the summation of all the permutations of the tensor products inside the bracket.
Therefore, a BSM can be performed in a concatenated manner: A lattice-level BSM ($\text{BSM}_\mathrm{lat}$) is done by $n$ block-level BSMs ($\text{BSM}_\mathrm{blc}$'s), each of which is again done by $m$ physical-level BSMs ($\text{BSM}_\mathrm{phy}$'s).
We refer to the sign (letter) result obtained from a lattice-, block-, or physical-level BSM as the \textit{lattice-, block-, or physical-level sign (letter)}, respectively.
\end{subequations}

\subsection{Original CBSM scheme}

We review the original CBSM scheme of the parity encoding in Ref.~\cite{lee2019fundamental}.
A $\text{BSM}_\mathrm{phy}$~can discriminate between only two among the four Bell states.
Three types of $\text{BSM}_\mathrm{phy}$'s ($B_\psi$, $B_+$, and $B_-$) are considered, which discriminate between $\qty{\ket{\psi^+}, \ket{\psi^-}}$, $\qty{\ket{\phi^+}, \ket{\psi^+}}$, and $\qty{\ket{\phi^-}, \ket{\psi^-}}$, respectively.
$B_\psi$ can be implemented by the process in Fig.~\ref{fig:physical_bsm_scheme}, which can be modified to implement $B_+$ instead by adding a \ang{45} wave plate on each input line just before the first PBS.
If the \ang{90} wave plate on the second input line is removed in the setting for $B_+$, $B_-$ is executed alternatively.
A $\text{BSM}_\mathrm{phy}$~has four possible outcomes: two successful cases (e.g., for $B_\psi$, $\ket{\psi^+}$ and $\ket{\psi^-}$), ``failure,'' and ``detecting a photon loss.''
Failure and loss can be distinguished by the number of total photons detected by the photon detectors.
Since two photons may enter a single detector, it is assumed that single-photon resolving detectors are used.
Note that, even in the failure cases, either sign or letter still can be determined.
(For example, even if a $B_\psi$ fails, we can still learn that the letter is $\phi$.)
On the other hand, if it detects a loss, we can get neither a sign nor a letter.

A $\text{BSM}_\mathrm{blc}$~is done by $m$-times of $\text{BSM}_\mathrm{phy}$'s.
Each block is composed of $m$ photons, thus we consider $m$ pairs of photons selected respectively in the two blocks.
The types of the $\text{BSM}_\mathrm{phy}$'s are selected as follows: First, $B_\psi$ is performed on each pair of photons in order until it either succeeds, detects a loss, or consecutively fails $j$ times, where $j \leq m-1$ is a predetermined number.
Then a sign $s=\pm$ is selected by the sign of the last $B_\psi$ outcome if it succeeds or selected randomly if it fails or detects a loss.
After that, $B_s$'s are performed for all the left pairs of photons.

The block-level sign (letter) is determined by the physical-level signs (letters) of the $m$ $\text{BSM}_\mathrm{phy}$'s.
In detail, the block-level sign is chosen (i) to be the same as $s$ if the last $B_\psi$ succeeds or any $B_s$ succeeds, and (ii) to be the opposite of $s$ if the last $B_\psi$ does not succeed and any $B_s$ fails.
(iii) Otherwise (namely, if the last $B_\psi$ does not succeed and all the $B_s$'s detect losses), the block-level sign is not determined.
The block-level letter is determined only when all the physical-level letters are determined, namely, when no losses are detected and all $B_s$'s succeed.
For such cases, the block-level letter is $\phi$ ($\psi$) if the number of $\psi$ in the $\text{BSM}_\mathrm{phy}$~results is even (odd).

Next, a $\text{BSM}_\mathrm{lat}$~is done by $n$-times of $\text{BSM}_\mathrm{blc}$'s.
The lattice-level sign is determined only when all the block-level signs are determined; it is $(+)$ if the number of $(-)$ in the $\text{BSM}_\mathrm{blc}$~results is even and it is $(-)$ if the number is odd.
The lattice-level letter is equal to any determined block-level letter.
Thus, if all $\text{BSM}_\mathrm{blc}$'s cannot determine letters, the lattice-level letter is not determined as well.

\subsection{Modified CBSM scheme for PTQC}

In our PTQC protocol, we consider using either single-photon resolving or on-off detectors.
The CBSM scheme should be slightly modified for this case.

Since failure and loss cannot be distinguished, a $\text{BSM}_\mathrm{phy}$~now has three possible outcomes: two successful cases and failure.
Consequently, in a $\text{BSM}_\mathrm{blc}$, $B_\psi$'s are performed until it either succeeds or consecutively fails $j$ times.
The way to determine the block-level sign and letter is the same as the original scheme, except that case (iii) when determining the sign no longer occurs.
The biggest difference from the original scheme is that the determined sign and letter may be wrong.
These error probabilities are presented in the next subsection.

In a $\text{BSM}_\mathrm{lat}$, the lattice-level sign is determined from the block-level signs by the same method as the original scheme, although it may be wrong with a nonzero probability as well.
On the other hand, the lattice-level letter is not determined by a single block-level letter unlike the original scheme; instead, we use a weighted majority vote of block-level letters.
The weight of each block-level letter is given as $w := \log\qty[ (1 - q_\mathrm{lett}^\mathrm{blc})/q_\mathrm{lett}^\mathrm{blc} ]$, where $q_\mathrm{lett}^\mathrm{blc}$ is the probability that the block-level letter is wrong.
This weight factor is justified as follows:
Let $I_\phi$ ($I_\psi$) denote the set of the indices of block pairs where the block-level letters are $\phi$ ($\psi$).
Assuming that the two lattice-level letters ($\Phi$ and $\Psi$) have the same prior probability, we get
\begin{align*}
    \frac{\Pr\qty(\Phi \middle| I_\phi, I_\psi)}{\Pr\qty( \Psi \middle| I_\phi, I_\psi)} &= \frac{\Pr\qty(I_\phi, I_\psi \middle| \Phi) \Pr\qty(\Phi)}{\Pr\qty(I_\phi, I_\psi \middle| \Psi) \Pr\qty(\Psi)} = \frac{\Pr\qty(I_\phi, I_\psi \middle| \Phi)}{\Pr\qty(I_\phi, I_\psi \middle| \Psi)} \\
    &= \frac{ \prod_{i \in I_\phi} \qty(1 - q_\mathrm{lett}^{(i)}) \prod_{i \in I_\psi} q_\mathrm{lett}^{(i)} }{ \prod_{i \in I_\phi} q_\mathrm{lett}^{(i)} \prod_{i \in I_\psi} \qty(1 - q_\mathrm{lett}^{(i)}) } \\
    &= \left. \prod_{i \in I_\phi} \frac{1 - q_\mathrm{lett}^{(i)}}{q_\mathrm{lett}^{(i)}} \middle/ \prod_{i \in I_\psi} \frac{1 - q_\mathrm{lett}^{(i)}}{q_\mathrm{lett}^{(i)}} \right. \\
    &= \exp\qty(\sum_{i=1}^n w^{(i)}), 
\end{align*}
where $q_\mathrm{lett}^{(i)}$ and $w^{(i)}$ are respectively the letter error probability and the weight of the $i$th block.
Note that the third equality comes from the fact that a lattice-level Bell state is decomposed into block-level Bell states of the same letter, as shown in Eqs.~\eqref{eq:logical_phi_decomposition} and \eqref{eq:logical_psi_decomposition}.

\subsection{Error probabilities of a CBSM under a lossy environment}

We here present the possible outcomes of a CBSM using either single-photon resolving or on-off detectors and the corresponding error probabilities $\qty(q_\mathrm{sign}, q_\mathrm{lett})$.
We denote $x := (1 - \eta)^2$, which is the probability that a $\text{BSM}_\mathrm{phy}$~does not detect photon losses.
It is assumed that the four Bell states have the same prior probabilities; namely, the initial marginal state on qubits 1 and 2 before suffering losses is the equal mixture of four lattice-level Bell states, which is justified in Supplementary Note~1.
For a $\text{BSM}_\mathrm{blc}$~or $\text{BSM}_\mathrm{lat}$, to avoid confusion, we use the term ``outcome'' to indicate the tuple of the outcomes of the $\text{BSM}_\mathrm{phy}$'s constituting the $\text{BSM}_\mathrm{blc}$~or $\text{BSM}_\mathrm{lat}$, and use the term ``result'' to indicate one of the four Bell states that gives the largest posterior probability under its outcome.
Note that the result of a BSM may be not deterministically determined by its outcome; if multiple Bell states have the same posterior probability, one of them is randomly selected as the result.

The case using single-photon resolving detectors is analyzed in Ref.~\cite{lee2019fundamental} and we here review the contents to be self-contained.
The outcome of a $\text{BSM}_\mathrm{blc}$~is included in one of the following three cases: (Success) Both the sign and letter are identified if no losses are detected and all the $B_\pm$'s succeed. (Failure) Neither sign nor letter is identified if no $B_\psi$'s succeed and all $B_\pm$'s detect losses. (Sign discrimination) Only the sign is identified if otherwise.
The block-level sign (or letter) is selected randomly if it is not identified.
The probabilities of these cases are respectively
\begin{equation*}
\begin{cases}
    \text{Success}: & p_\mathrm{s} = \qty[1 - 2^{-(j+1)}] x^m, \\
    \text{Failure}: & p_\mathrm{f} = \sum_{l=0}^j \qty(\frac{x}{2})^l \qty(1 - x)^{m-l}, \\
    \text{Sign discrimination}: & p_\mathrm{sd} = 1 - p_\mathrm{s} - p_\mathrm{f}.
\end{cases}
\end{equation*}
For a $\text{BSM}_\mathrm{lat}$, let $N_\mathrm{s}$ ($N_\mathrm{f}$) denote the number of successful (failed) $\text{BSM}_\mathrm{blc}$'s.
The lattice-level letter is identified if $N_\mathrm{s} \geq 1$ (namely, if at least one block-level letter is identified) and the sign is identified if $N_\mathrm{f} = 0$ (namely, if all block-level signs are identified).
Hence, the outcome of a $\text{BSM}_\mathrm{lat}$~is included in one of the following four events:
\begin{align*}
    \begin{cases}
    S~(\text{Success}): & N_\mathrm{s} \geq 1 \land  N_\mathrm{f} = 0, \\
    D_L~(\text{Letter discrimination}): & N_\mathrm{s}, N_\mathrm{f} \geq 1, \\
    D_S~(\text{Sign discrimination}): & N_\mathrm{s} = N_\mathrm{f} = 0, \\
    F~(\text{Failure}): & N_\mathrm{s}=0 \land N_\mathrm{f} \geq 1.
    \end{cases}
\end{align*}
The sign and letter error probabilities $(q_\mathrm{sign}, q_\mathrm{lett})$ of the $\text{BSM}_\mathrm{lat}$~for each event are $(0, 0)$ for $S$, $(1/2, 0)$ for $D_L$, $(0, 1/2)$ for $D_S$, and $(1/2, 1/2)$ for $F$.
The probabilities of the events are respectively given as
\begin{align}
\begin{split}
    P_S &= \qty(1 - p_\mathrm{f})^n - p_\mathrm{sd}^n, \\
    P_{D_L} &= 1 - \qty(1 - p_\mathrm{s})^n + \qty(1 - p_\mathrm{f})^n - p_\mathrm{sd}^n, \\
    P_{D_S} &= p_\mathrm{sd}^n, \\
    P_F &= \qty(1 - p_\mathrm{s})^n - p_\mathrm{sd}^n.
\end{split}
\label{eq:CBSM_probability_sprd}
\end{align}


We now consider using on-off detectors for fusions.
Each outcome of a $\text{BSM}_\mathrm{blc}$~is uniquely identified by a triple $O = \qty(r,s,\vb{U})$, where $r \in \mathbb{Z}_{j+1} := \qty{0, \cdots, j}$ is the number of failed $B_\psi$'s, $s = \pm$ is the sign chosen by the successful $(r+1)$th $B_\psi$ (if $r < j$) or randomly (if $r = j$), and $\vb{U}$ is an $\qty(m - r)$-element tuple composed of ``$\phi$,'' ``$\psi$,'' and ``$f$'' (failure) indicating the outcomes of the $\text{BSM}_\mathrm{phy}$'s from the $(r+1)$th to the the last.
(If $r < j$, the first component of $\vb{U}$ is always $\psi$, and the other components are determined by the $B_s$'s. If $r = j$, all the components are determined by the $B_s$'s.)
Let $N_e(\vb{U})$ for $e \in \qty{\phi, \psi, f}$ denote the number of $e$ in $\vb{U}$.
Then a $\text{BSM}_\mathrm{blc}$~outcome $O$ is included in one of the following $j+3$ events:
\begin{align}
\begin{split}
    \mathcal{S}_r &:= \qty{(r, s, \vb{U}) \middle| N_f(\vb{U}) = 0} \quad (0 \leq r \leq j), \\
    \mathcal{F} &:= \qty{(j, s, \vb{U}) \middle| N_f(\vb{U}) = m - j}, \\
    \mathcal{D} &:= \mathcal{O} \setminus \qty[ \mathcal{F} \cup \bigcup_{r=0}^j \mathcal{S}_r ],
\end{split}
\label{eq:bsm1_events}
\end{align}
where $\mathcal{O}$ is the set of all possible outcomes.
Note that the events $\mathcal{S}_r$, $\mathcal{F}$, and $\mathcal{D}$ correspond to success, failure, and sign discrimination when $\eta = 0$.
For each event $\mathcal{E}$ in Eq.~\eqref{eq:bsm1_events}, its sign and letter error probabilities $q_\mathrm{sign/lett}^\mathrm{blc}(\mathcal{E})$ and the probability $p_\mathcal{E}$ that the event occurs are given as follows (see Supplementary Note~6 for their derivation):
\begin{align}
    \begin{split}
    &\begin{cases}
        q_\mathrm{sign}^\mathrm{blc}\qty(\mathcal{S}_r) = 0, \qquad q_\mathrm{lett}^\mathrm{blc}\qty(\mathcal{S}_r) = \frac{1}{2} - \frac{1}{2} \qty( \frac{x}{2-x} )^r, \\
        p_{\mathcal{S}_r} = \frac{1}{2} \qty(1 - \frac{x}{2})^r x^{m-r},
    \end{cases} \\
    &\begin{cases}
        q_\mathrm{sign}^\mathrm{blc}\qty(\mathcal{F}) = \frac{(1 - x)^{m-j}}{ 1 + (1 - x)^{m-j}}, \qquad q_\mathrm{lett}^\mathrm{blc}\qty(\mathcal{F}) = \frac{1}{2}, \\
        p_{\mathcal{F}} = \frac{1}{2} \qty(1 - \frac{x}{2})^j \qty[ 1 + (1 - x)^{m-j} ],
    \end{cases} \\
    &\begin{cases}
        q_\mathrm{sign}^\mathrm{blc}\qty(\mathcal{D}) = 0, \qquad q_\mathrm{lett}^\mathrm{blc}\qty(\mathcal{D}) = \frac{1}{2}, \\
        p_{\mathcal{F}} = 1 - \sum_r p_{\mathcal{S}_r} - p_\mathcal{F}.
    \end{cases}
    \end{split}
    \label{eq:CBSM_block_error_prob_onoff}
\end{align}
A possible outcome of a $\text{BSM}_\mathrm{lat}$ corresponds to an $n$-tuple of events composed of $\mathcal{S}_r$ ($0 \leq r \leq j$), $\mathcal{F}$, and $\mathcal{D}$, which can be regarded as an independent event for the outcomes of the $\text{BSM}_\mathrm{lat}$.
The probability that an event $\mathbf{E} = \qty(\mathcal{E}_1, \cdots, \mathcal{E}_n)$ occurs is
\begin{align}
    p_\mathbf{E} &= \prod_{i=1}^n p_{\mathcal{E}_i}
    \label{eq:CBSM_probability_onoff}
\end{align}
and the sign and letter error probabilities of $\mathbf{E} = \qty(\mathcal{E}_1, \cdots, \mathcal{E}_n)$ are respectively
\begin{align*}
    q_\mathrm{sign}(\mathbf{E}) &= \frac{1}{2} - \frac{1}{2} \qty[1 - 2q_\mathrm{sign}^\mathrm{blc}(\mathcal{F})]^{N_\mathcal{F}},\\
    q_\mathrm{lett}(\mathbf{E}) &= \frac{1}{2} + \frac{1}{2} \sum_{(\lambda_1, \cdots, \lambda_n) \in \mathbb{Z}_2^n} \prod_{i=1}^n \qty[ q_i^{\lambda_i} \qty(1 - q_i)^{1 - \lambda_i} ] \\
    &\qquad\qquad \times \mathrm{sgn}\qty( \sum_{i=1}^n \qty(2\lambda_i - 1) \log\frac{1 - q_i}{q_i} ),
\end{align*}
where $N_\mathcal{F}$ is the number of $\mathcal{F}$'s in $\mathbf{E}$, $q_i := q^\mathrm{blc}_\mathrm{lett}\qty(\mathcal{E}_i)$, and $\mathrm{sgn}(a)$ is $a/\abs{a}$ if $a \neq 0$ and 0 if $a = 0$.
See Supplementary Note~6 for their derivation.

\subsection{Generation of post-$H$ microclusters}

\begin{figure*}[t!]
	\centering
	\includegraphics[width=\textwidth]{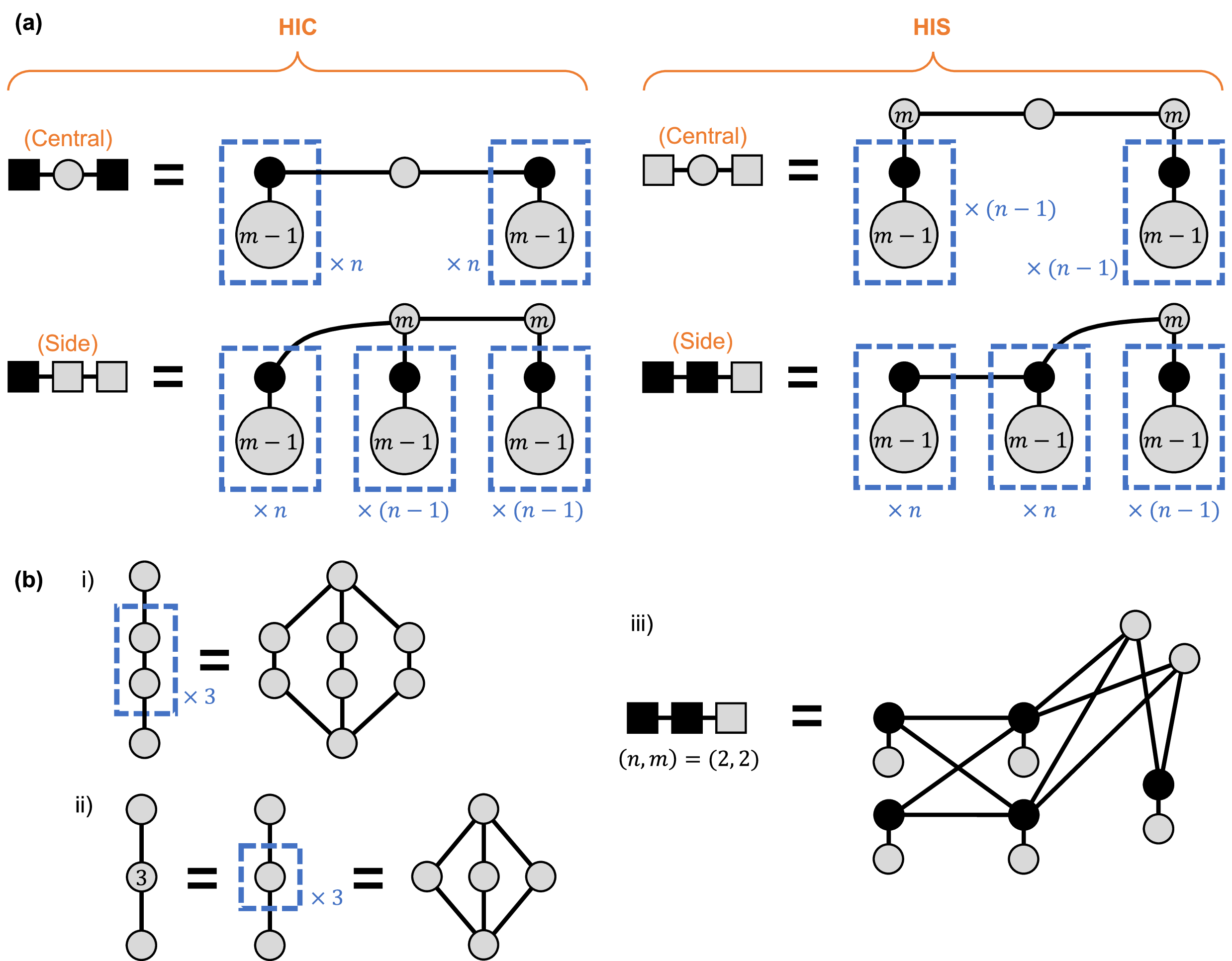}
	\caption{
	    \textbf{Physical-level graphs of post-$H$ microclusters for PTQC.}
        In (a), the physical-level graphs are shown for the HIC and HIS configurations when the $(n, m)$ parity encoding is used for PTQC.
	    The squares (circles) correspond to lattice-level (physical-level) qubits, among which black ones indicate that the lattice-level (physical-level) Hadamard gates are applied to the qubits on the graph state.
	    A blue dashed box indicates a group of recurrent subgraphs; that is, the structure in the box is repeated as many times as indicated, and if there is an edge across the border of the box, it means that edges of the same pattern exist in each of the repeated structures.
	    See i) of (b) for an example.
	    A number inside a circle means a blue dashed box surrounding only the circle with the indicated repetition number, as shown in the example of ii) of (b).
	    If there is an edge between two blue dashed boxes or circles containing numbers, the full graph can be recovered just by expanding them one by one.
	    As an example, the full graph of the side microcluster of the HIS configuration for $n=m=2$ is shown in iii) of (b).
	}
	\label{fig:microcluster_structures}
\end{figure*}

In this subsection, we first present the physical-level graphs of post-$H$ microclusters for PTQC and then describe the method to generate them.
A post-$H$ microcluster, which is composed of three lattice-level qubits or two of them and one photon (physical-level qubit), can be regarded as a graph state of photons up to several physical-level Hadamard gates.
The graph of this graph state, called the \textit{physical-level graph} of the post-$H$ microcluster, is visualized in Fig.~\ref{fig:microcluster_structures} for each post-$H$ microcluster; see Supplementary Note~7 for their derivation.
Here, the squares (circles) indicate lattice-level (physical-level) qubits.
If a square (circle) is filled with black, it means that the lattice-level (physical-level) Hadamard gate is applied on the qubit after the involved edges are connected.
Recurrent subgraphs are abbreviated as blue dashed squares or circles with numbers; see Fig.~\ref{fig:microcluster_structures}(b) for the detailed interpretation of these notations.

We now depict the ways to generate a specific post-$H$ microcluster from GHZ-3 states.
We first describe a straightforward method and then adjust or generalize it.
The final method can be summarized as follows:
\begin{enumerate}
    \item Determine a \textit{merging graph} $G$ for the post-$H$ microcluster that we want to create by the algorithm presented below. Each edge of $G$ is labeled as either ``internal" or ``external."
    \item For each vertex $v$ in $G$, Prepare a GHZ-3 state $\ket{\mathrm{GHZ}_3}_v$.
    \item For each edge $e$ in $G$ that connects $v_1$ and $v_2$, perform a BSM (fusion) on two photons selected respectively from $\ket{\mathrm{GHZ}_3}_{v_1}$ and $\ket{\mathrm{GHZ}_3}_{v_2}$ if $e$ is an internal (external) edge. The order of the operations does not matter.
\end{enumerate}

We define the \textit{GHZ-$l$ state} for an integer $l \geq 3$ by the state $\ket{\mathrm{GHZ}_l} := \ket{\textsc{h}}^{\otimes l} + \ket{\textsc{v}}^{\otimes l}$.
Note that it is a state obtained from a graph state with a star graph (where the number of vertices is $l$) by applying Hadamard gates on all the leaves of the graph; namely,
\begin{align*}
    \ket{\mathrm{GHZ}_l} = H_2 \cdots H_l C^Z_{12} \cdots C^Z_{1l} \ket{+}^{\otimes l}.
\end{align*}
We refer to the first photon of the above expression as the \textit{root photon} of the state (which can be chosen arbitrarily) and the other photons as its \textit{leaf photons}.

\begin{figure}[ht!]
	\centering
	\includegraphics[width=\columnwidth]{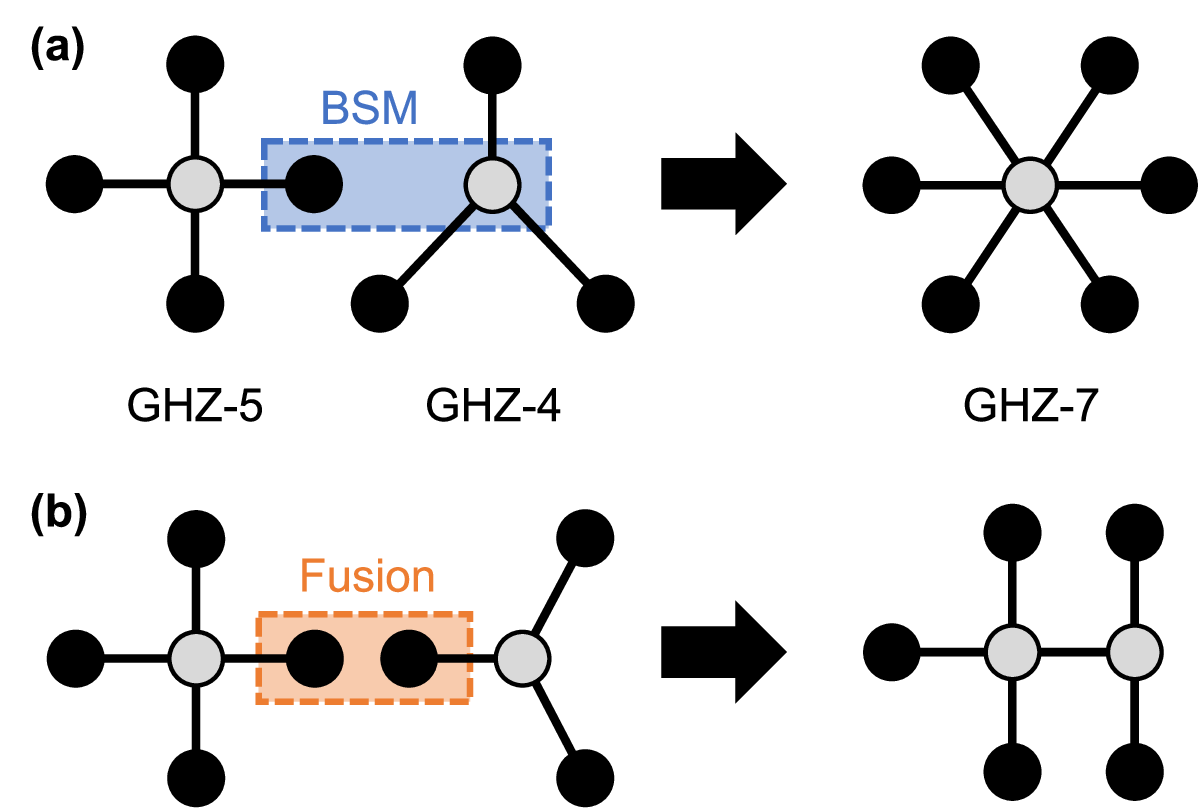}
	\caption{
	    \textbf{Examples of the two types of merging operations on two GHZ states.}
        (a) A BSM on the root photon of one state and a leaf photon of the other and (b) a fusion on two leaf photons are shown.
	}
	\label{fig:merging_GHZs}
\end{figure}

If a BSM is performed on the root photon of a GHZ-$l_1$ state and a leaf photon of a GHZ-$l_2$ state, the resulting state on the remaining photons is a GHZ-$(l_1 + l_2 - 2)$ state; see Fig.~\ref{fig:merging_GHZs}(a) for an example.
Thus, an arbitrary GHZ state can be constructed by performing BSMs on multiple GHZ-3 states appropriately.
On the other hand, if a \textit{fusion} is performed on two leaf photons selected respectively from GHZ-$l_1$ and GHZ-$l_2$ states, the resulting state is no longer a GHZ state, but it is a graph state (up to some Hadamard gates) with a graph containing a vertex with degree $l_1 - 1$, a vertex with degree $l_2 - 1$, and multiple vertices with degree one; see Fig.~\ref{fig:merging_GHZs}(b) for an example.
(The degree $d_v$ of a vertex $v$ means the number of edges connected to $v$.)

Combining the above facts, a post-$H$ microcluster (or an arbitrary graph state) with the physical-level graph $G$ can be generated from GHZ-3 states up to physical-level Hadamard gates in the following way: For each vertex $v$ of $G$ with a degree larger than one, prepare a state $\ket{\mathrm{GHZ}_{d_v+1}}_v$ through BSMs on GHZ-3 states.
Then, for each edge $\qty(v_1, v_2)$ of $G$, perform a fusion on two photons selected respectively from $\ket{\mathrm{GHZ}_{d_{v_1}+1}}_{v_1}$ and $\ket{\mathrm{GHZ}_{d_{v_2}+1}}_{v_2}$.
We refer to each BSM or fusion during this process as a \textit{merging operation}.

\begin{figure}[t!]
	\centering
	\includegraphics[width=\columnwidth]{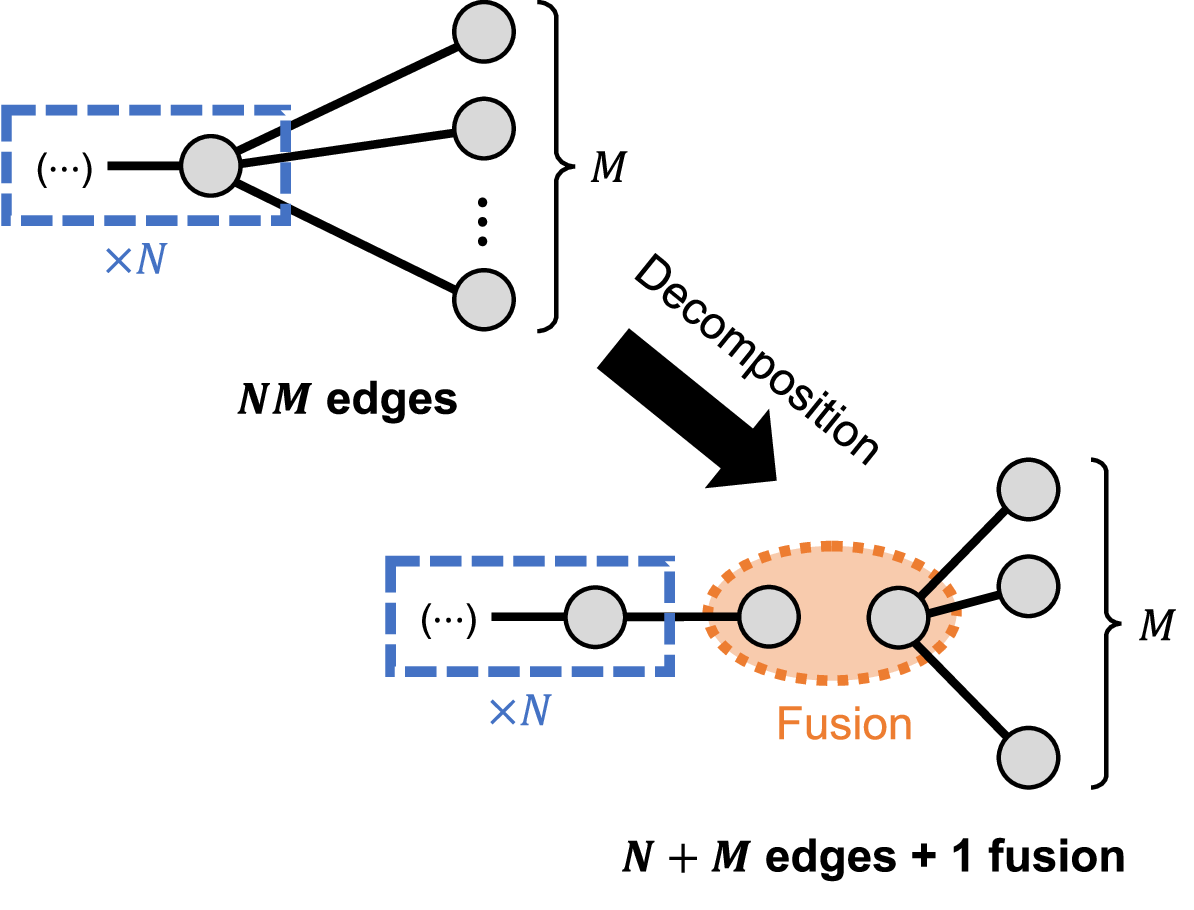}
	\caption{
	    \textbf{Decomposition of a graph state.}
        A graph state is decomposed by separating recurrent subgraphs that are connected with multiple vertices.
	}
	\label{fig:decomp_process}
\end{figure}

However, the above method still has room for improvement.
The physical-level graphs in Fig.~\ref{fig:microcluster_structures} can be decomposed into multiple components that are combined by fusions through the process shown in Fig.~\ref{fig:decomp_process}.
Here, each recurrent subgraph connected with multiple vertices is separated and connected with only one vertex.
The decomposition of different post-$H$ microclusters is explicitly presented in Supplementary Figure~8.
To generate a post-$H$ microcluster, we prepare the individual components first by the aforementioned method, then merge them through fusions.
This process may greatly reduce the number of required merging operations since the number of edges decreases as shown in Fig.~\ref{fig:decomp_process}.

\begin{figure*}[t!]
	\centering
	\includegraphics[width=0.91525424\textwidth]{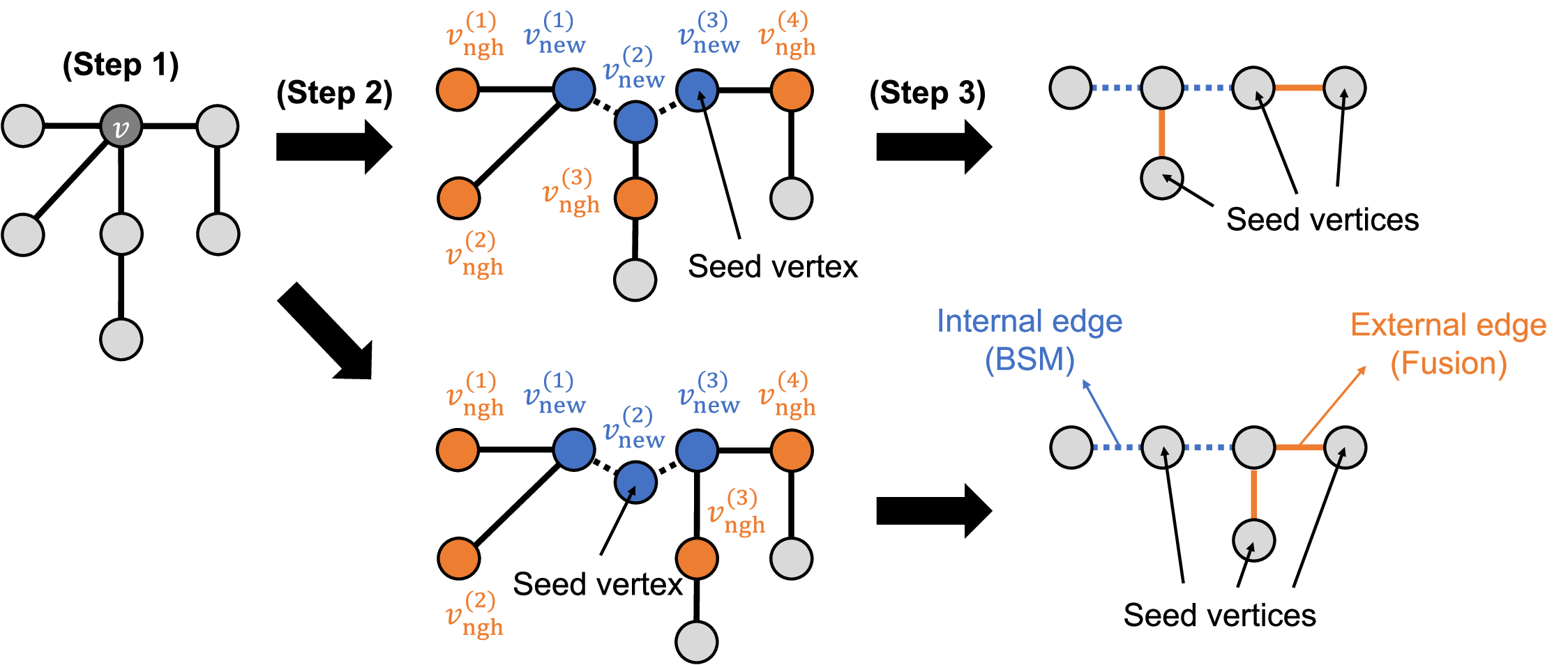}
	\caption{
	    \textbf{Construction of merging graphs from a physical-level graph.}
	    $v$ is the only vertex with a degree larger than two in the original graph.
	    The upper and lower processes differ in the selection of the seed vertex for the decomposition of $v$.
	}
	\label{fig:merging_graph}
\end{figure*}

Furthermore, we can generalize the method using the fact that every merging operation commutes with each other.
That is, even if all the fusions and BSMs in the above process are performed in an arbitrary order, the final state does not vary (up to the change of the Pauli frame).
To systematically address this feature, we define a \textit{merging graph} of a post-$H$ microcluster or one of its components by a graph in which the vertices correspond to initial GHZ-3 states and the edges indicate the merging operations between them required to generate the state.
Each edge of a merging graph is either \textit{internal} or \textit{external} that corresponds to BSMs or fusions, respectively.

A merging graph of a component can be constructed by the following method starting from its physical-level graph (see Fig.~\ref{fig:merging_graph} for two examples):
First, for each vertex $v$ satisfying $d_v \geq 2$ in the physical-level graph, replace it with $d_v - 1$ new vertices connected by internal edges in series with each other.
This process means decomposing a GHZ-$\qty(d_v+1)$ state into $d_v - 1$ GHZ-3 states.
The edges originally connected to $v$ are distributed to the new vertices in a way that every new vertex is connected to three or fewer edges.
Then there is only one vertex connected to two edges, which is called the \textit{seed} vertex of $v$.
This seed vertex means that one photon in the corresponding GHZ-3 state does not participate in any merging operation and remains in the final state.
Lastly, the merging graph is obtained by removing all the vertices with degree one.
See Supplementary Note~8 for a stricter step-by-step description of the method.

The merging graph of a post-$H$ microcluster is constructed by combining the merging graphs of its components.
That is, for each fusion between different components, the corresponding seed vertices in the merging graphs are connected by an external edge.
Then we finally get the method summarized at the beginning of this subsection.

\subsection{Optimization of resource overheads}

The process of generating a post-$H$ microcluster described above is determined by two factors: the merging graph and the order of the merging operations.
Here, we discuss their optimization for minimizing resource overhead.
The merging graph is selected randomly by the algorithm in Supplementary Note~8.
Based on it, we determine the order of the merging operations through an algorithm found heuristically and calculate the expected number $N_\mathrm{GHZ}^{\mathrm{MC}}$ of GHZ-3 states required to generate the state.
We repeat this process for a large enough number to obtain as low resource overhead as possible.
$N_\mathrm{GHZ}^*$ and $\mathcal{N}_{p_L}$ can be calculated by using the obtained optimal resource overheads; see Supplementary Note~3 for details.

During the generation process, performing each merging operation can be regarded as contracting the corresponding edge, which means removing the edge, merging the two vertices $\qty(v_1, v_2)$ that it previously joined into a new vertex $w$, and reconnecting all the edges that were connected to $v_1$ and $v_2$ with $w$.
Here, each vertex indicates a connected subgraph (a group of entangled photons) of the intermediate graph state.
We assign a ``weight'' $N_v$ (which is initialized to 1) on each vertex $v$, which is the average number of GHZ-3 states required to generate the connected subgraph.
If the edge between two vertices $v_1$ and $v_2$ are contracted, the new vertex $w$ has the weight of
\begin{align}
    N_w = \frac{2}{(1 - \eta)^2} \qty(N_{v_1} + N_{v_2}) =: N_{v_1} +_m N_{v_2},
    \label{eq:fusion_sum_def}
\end{align}
where the factor $2/(1 - \eta)^2$ is the inverse of the success probability of the merging operation.
By repeating this process, the post-$H$ microcluster is obtained when there is only one vertex left, whose weight is equal to $N_\mathrm{GHZ}^{\mathrm{MC}}$.

To find an optimal order of merging operations, we use the following strategy:
\begin{enumerate}
    \item Find the set $E_\mathrm{min.wgt}$ of edges with the smallest weight, where the weight of an edge $\qty(v_1, v_2)$ is defined as $N_{v_1} +_m N_{v_2}$.
    \item Using an edge coloring algorithm, allocate ``colors'' to all edges so that different edges sharing a vertex have different colors and as few colors as possible are used.
    \item Partition $E_\mathrm{min.wgt}$ into disjoint subsets by the colors of the edges.
    Find the largest subset $E_\mathrm{mrg}$ among them. If such a subset is not unique, choose one randomly.
    \item Contract each edge in $E_\mathrm{mrg}$ in an arbitrary order.
    \item Repeat all the above steps until only one vertex is left.
\end{enumerate}
The strategy is based on the following two intuitions:
First, it is better to merge vertices with small weights first, since $(N_1 +_m N_2) +_m N_3 < N_1 +_m (N_2 +_m N_3)$ if $N_1 < N_2 < N_3$.
Secondly, it is better to perform merging operations in parallel as much as possible.
Such a set of edges can be found by the edge coloring algorithm.
For our results, we have used the function \texttt{coloring.greedy\_color} in NetworkX package \cite{aric2008exploring} with the strategy \texttt{largest\_first}.
(Since the function performs vertex coloring, we input the line graph of $G_\mathrm{mrg}$ into the function.)

In Supplementary Note~9, we show an evidence that this optimizing strategy is indeed highly effective in terms of both the optimality of the calculated overhead and searching time, by comparing its performance with those of its variants constructed by omitting or altering specific steps.
We conjecture that this strategy is powerful for generating general graph states as well as those for PTQC, which will be worth investigating.

\subsection{Conversion of resource measures}

We here address the conversion of resource measures for a fair comparison between PTQC and the protocol in Ref.~\cite{li2015resource}.
In Ref.~\cite{li2015resource}, resource overheads are quantified by the number of photodetectors required per central qubit, not the number of GHZ-3 states we have used, thus conversion between them is necessary for a fair comparison.
In PTQC, detectors are used when generating GHZ-3 states, applying physical-level BSMs, and measuring central qubits.
We suppose that GHZ-3 states are generated by the scheme proposed in Ref.~\cite{varnava2008how} like the protocol in Ref.~\cite{li2015resource}.
The scheme uses six detectors to generate a single GHZ-3 state and succeeds with probability $1/32$; thus, generating one GHZ-3 state requires 192 detectors.
(If it is allowed to use photodetectors repeatedly during the generation of each GHZ-3 state, only six detectors are required per GHZ-3 state. However, we ignore this option to be consistent with Ref.~\cite{li2015resource}.)
Next, four detectors are used for one physical-level BSM (see Fig.~\ref{fig:physical_bsm_scheme}).
Counting the number of physical-level BSMs per central qubit is not simple, but we can get its upper bound as $(3N_\mathrm{GHZ}^* - 1)/2$, which is half the number of total photons in all GHZ-3 states except one photon in the central qubit.
Lastly, two detectors are used for the two polarization modes when measuring a central qubit.
In total, $N_\mathrm{det} = 198N_\mathrm{GHZ}^*$ detectors are required per data qubit in PTQC.
Since $N_\mathrm{GHZ}^* \gtrapprox 330$ is required for a positive photon loss threshold (see Fig.~\ref{fig:thres_vs_overheads}), $N_\mathrm{det}$ should be at least about $7 \times 10^4$.

\section*{Data Availability}

All the numerical data used to generate the figures are available from the corresponding author upon reasonable request.

\section*{Code Availability}

The Python codes used for numerical simulations are available from the corresponding author upon reasonable request.

\section*{Author Contributions}

All authors (S.H.L., S.O., Y.S.T., and H.J.) contributed to developing the main idea. S.H.L. concretized the idea with mathematical analysis, wrote the codes, and ran numerical simulations. S.O. suggested important ideas on resource analysis. Y.S.T. checked the results and H.J. supervised the project. All authors helped write the manuscript.

\section*{Competing Interests}

The authors declare no competing interests.

\section*{Acknowledgements}

This work was supported by the National Research Foundation of Korea (NRF) grants funded by the Korean government (Grant~Nos.~NRF-2020R1A2C1008609,~NRF-2019R1A6A1A10073437, NRF-2022M3E4A1076099, and 2022M3K4A1097117) via the Institute of Applied Physics at Seoul National University, and by the Institute of Information \& Communications Technology Planning \& Evaluation (IITP) grant funded by the Korea government (MSIT) (IITP-2021-0-01059 and IITP-2022-2020-0-01606).
We thank Kamil Bradler, Brendan Pankovich, Angus Kan, and Alex Neville for insightful discussions.

\bibliography{references}

\end{document}


\title{\textit{Supplemental Material} \\ Parity-encoding-based quantum computing with Bayesian error tracking}

\author{Seok-Hyung Lee}
\affiliation{Department of Physics and Astronomy, Seoul National University, Seoul 08826, Republic of Korea}
\author{Srikrishna Omkar}
\affiliation{ORCA Computing, Toronto M6P3T1, Canada}
\author{Yong Siah Teo}
\affiliation{Department of Physics and Astronomy, Seoul National University, Seoul 08826, Republic of Korea}
\author{Hyunseok Jeong}
\affiliation{Department of Physics and Astronomy, Seoul National University, Seoul 08826, Republic of Korea}

\maketitle

\section*{Supplementary Note~1: Proof of the maximally-mixedness of a marginal state in a graph state}
\label{SN:marginal_state}

We here verify that the marginal state on the qubits participating in a fusion is maximally mixed.
More strictly, we prove the statement: \textit{For a graph state $\ket{G}_V$ with a graph $G = (V, E)$ and given two vertices $a, b \in V$, if $\qty{a} \cup N\qty(a)$ and $\qty{b} \cup N\qty(b)$ are disjoint and neither $N(a)$ nor $N(b)$ is empty where $N(v)$ for a vertex $v \in V$ is the set of vertices adjacent to $v$, the marginal state $\Tr_{V\setminus\qty{a, b}} \ketbra{G}_V =: \rho_{ab}$ is maximally mixed.}

Let $\mathcal{S}$ denote the stabilizer group of the zero-dimensional Hilbert space $\qty{\ket{G}_V}$.
First, any stabilizer $S \in \mathcal{S}$ can be written as the product of stabilizer generators: $S = \prod_{v \in V_0} S_v$ where $V_0 \subseteq V$ and $S_v := X_v \prod_{v' \in N(v)} Z_{v'}$.
If $V_0$ contains a vertex $c \neq a, b$, $S$ must contain $X_c$ or $Y_c$ since no stabilizer generators besides $S_c$ contain $X_c$.
If otherwise, $V_0$ is one of $\emptyset$, $\qty{a}$, $\qty{b}$, and $\qty{a, b}$.
Except when $V_0$ is empty (namely, $S$ is identity), there exists a vertex $c \neq a, b$ such that $S$ contains $Z_c$, since $N(a)$ and $N(b)$ are not empty, $b \notin N(a)$, $a \notin N(b)$, and $N(a) \neq N(b)$.
Therefore, every single- or two-qubit Pauli operator on $a$ and $b$ that is not identity cannot be a stabilizer, thus it anticommutes with at least one stabilizer.
(If such an operator $P_a P_b$ commutes with all stabilizers, $P_a P_b\ket{G}_V$ is also stabilized by $\mathcal{S}$, which means that $P_a P_b\ket{G}_V = \ket{G}_V$ since $\mathcal{S}$ stabilizes the zero-dimensional Hilbert space.)
Consequently, $\Tr( P_a P_b \rho_{ab}) = \bra{G} P_a P_b \ket{G} = 0$ for every single- or two-qubit Pauli operator $P_a P_b$ that is not identity.
The state $\rho_{ab}$ satisfying this condition is unique and maximally mixed.

\newpage

\section*{Supplementary Note~2: Details of error simulations}
\label{SN:error_simulations}

Here, we describe the error simulation method in detail.
We first introduce the parameters that determine the details of PTQC:
\begin{itemize}
    \item \texttt{pssl}: If \true, star clusters generated by successful step-1 fusions are post-selected for step 2. If \false, all generated star clusters are used regardless of the fusion results.
    \item \texttt{hic}: If \true, the $H$-configuration is HIC. If \false, it is HIS.
    \item \texttt{sprd}: If \true, single-photon resolving detectors are used. If \false, on-off detectors are used.
    \item $n$, $m$: The $(n, m)$ parity encoding is used to encode side qubits.
    \item $j$: The maximal number of $B_\psi$'s in a \bsmb.  (See the CBSM scheme in the Methods section of the main text.)
\end{itemize}

For a fixed parameter setting, we consider an RHG lattice whose boundaries are in the form of a cuboid as visualized in Supplementary Figure~\ref{fig:identity_gate}, which implements a logical identity gate.
Let us term the three axes of the cuboid as the $x$-, $y$-, and $t$-axis and the corresponding boundaries as the $x$-, $y$-, and $t$-boundaries.
The $t$-axis is also referred to as the \textit{simulated time axis}.
The cuboid has the widths of $d-1$ unit cells along the $x$- and $y$-axis, where $d$ is the code distance, and the width of $T = 4d + 1$ unit cells along the $t$-axis.
The value of $T$ is arbitrarily set to be larger enough than $d$ for reducing the effects of errors near the $t$-boundaries.
The $x$- and $t$-boundaries are set to be primal, while the $y$-boundaries are set to be dual.
In other words, the $x$- and $t$-boundaries adjoin normally on primal unit cells, while the $y$-boundaries cross the middle of primal unit cells.
For error simulations, we count error chains connecting the opposite $x$-boundaries, thus we assume that the qubits on the $t$-boundaries do not have errors.

\begin{figure}[ht!]
    \centering
    \includegraphics[width=0.77482014\textwidth]{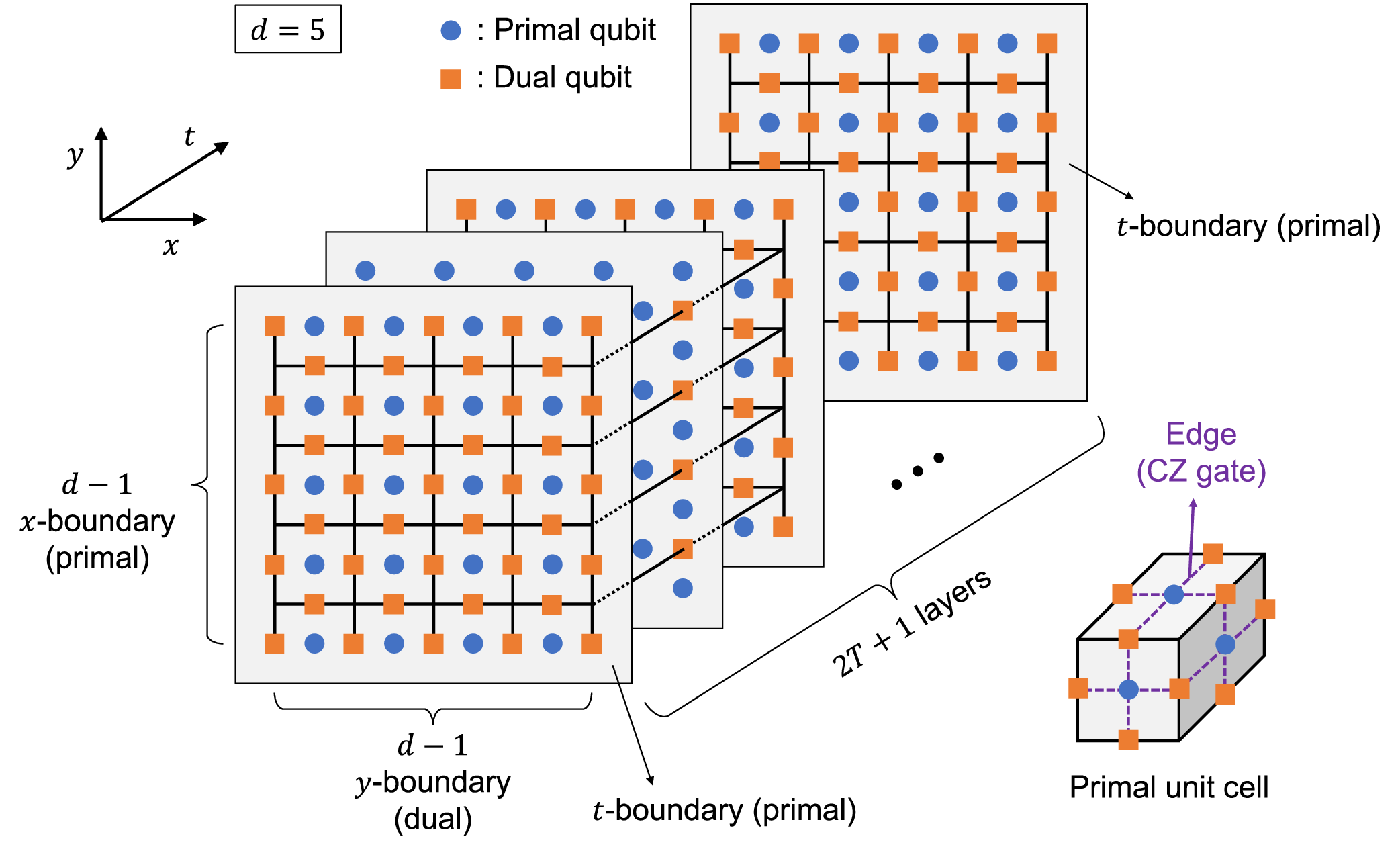}
    \caption{
        \textbf{Structure of a logical identity gate for simulations.}
        The code distance is $d = 5$ and the length along the simulated time ($t$) axis is $T$ in the unit of a cell.
    }
    \label{fig:identity_gate}
\end{figure}

We use a Monte-Carlo method for the simulations.
Each trial is proceeded as follows:
\begin{enumerate}
    \item Sample the outcomes of all fusions in steps 1 and 2 (only step 2 if \texttt{pssl} is \true) by the probabilities shown in the Methods section of the main text, which depend on the values of $n$, $m$, $j$, and $\eta$.
    \item For each fusion outcome, the corresponding error probabilities ($q_\mr{sign}$, $q_\mr{lett}$) are obtained and whether the fusion has a sign or letter error is randomly determined by the probabilities.
        These error probabilities and errors are then propagated to appropriate central qubits determined by the value of \texttt{hic}.
        For each central qubit~$i$, the presence or absence of an error and its probability are assigned to a boolean variable $\mathtt{error}_i$ and a floating-point variable $q_{\mr{err},i}$, respectively.
    \item For each central qubit~$i$, a photon loss is sampled with probability $\eta$. 
        If it has a loss, $q_{\mr{err},i}$ is updated to $0.5$ and $\mathtt{error}_i$ is flipped with probability 50\%.
    \item The syndrome of each parity-check operator (which corresponds to a primal unit cell) is determined by the values of $\mathtt{error}_i$'s of the qubits in the support of the operator.
    \item The syndromes are decoded to infer the locations of the errors.
    We use the weighted minimum-weight perfect matching decoder via PyMatching package \cite{higgott2021pymatching} where the weight for each qubit~$i$ is $\log[\qty(1 - q_{\mr{err},i}) / q_{\mr{err},i}]$.
    (If $q_{\mr{err},i} = 0$, the weight is infinity, which is handled by ignoring the qubit from the input of the decoder.)
    Exceptionally, if every value of $q_{\mr{err},i}$ is either $0$ or $1/2$, the qubits with $q_{\mr{err},i} = 1/2$ are given the weight of one, not zero, for a technical reason.
    \item The remaining errors are obtained by comparing the original and estimated errors.
    If the number of the remaining errors on one side of the $x$-boundaries is odd, we regard that this trial has a logical error.
\end{enumerate}

The logical error rate $p_L$ for a given parameter setting is obtained by repeating the above process a sufficient number of times.
In detail, we repeat the process until $\Delta p_L / p_L \leq 0.1$ is reached where $\Delta p_L$ is half the width of the 99\% confidence interval.
The logical error rates $p_L^{(9)}(\eta)$, $p_L^{(11)}(\eta)$ are calculated while varying $\eta$ for two code distances $d = 9,~11$ and the loss threshold $\eta_\mr{th}$ is obtained by finding the largest $\eta$ satisfying $p_L^{(11)}(\eta) + \Delta p_L^{(11)}(\eta) < p_L^{(9)}(\eta) - \Delta p_L^{(9)}(\eta)$.

\newpage

\section*{Supplementary Note~3: Details of resource analysis}
\label{SN:resource_analysis}

Here, we describe the details of resource analysis on PTQC.
We first investigate calculating $N_\mr{GHZ}^*$, the expected number of required GHZ-3 states to generate one star cluster.
$N_\mr{GHZ}^*$ is used to obtain the expected total number $\mathcal{N}_{p_L^\mr{targ}}$ of GHZ-3 states to achieve the target logical error rate of $p_L^\mr{targ}$ for the logical identity gate with the length of $d-1$ unit cells.

By using the optimization method presented in the Methods section of the main text, we determine the merging graphs and the orders of the merging operations for center and side post-$H$ microclusters and calculate their resource overheads $N_\mr{GHZ}^\mr{central}$ and $N_\mr{GHZ}^\mr{side}$.
Then we get
\begin{align*}
    N^*_\mr{GHZ} = \begin{cases}
        \qty[\qty(N_\mr{GHZ}^\mr{central} + N_\mr{GHZ}^\mr{side}) / p_\mr{succ,step1} + N_\mr{GHZ}^\mr{side}]/p_\mr{succ,step1} & \text{if \texttt{pssl} is \true}, \\
        N_\mr{GHZ}^\mr{central} + 2N_\mr{GHZ}^\mr{side} & \text{if \texttt{pssl} is \false}, \\
    \end{cases}
\end{align*}
where $p_\mr{succ,step1}$ is the average success probability of step-1 fusions and \texttt{pssl} is defined in Supplementary Note 2.

To obtain the simulation results in the main text, we sample 1200 values of $N_\mr{GHZ}^\mr{MC}$ through the aforementioned process.
Let $N_1$ ($N_2$) be the minimal values of $N_\mr{GHZ}^\mr{MC}$ for the first 600 (total 1200) samples.
If $N_1 = N_2$, the value is returned.
If otherwise, we sample 1200 values of $N_\mr{GHZ}^\mr{MC}$ again and denote the minimal $N_\mr{GHZ}^\mr{MC}$ for the total 2400 samples by $N_3$.
If $N_2 = N_3$, the value is returned.
If otherwise, we sample 2400 merging graphs again and so on.
By varying the total number of samples in this way, it is possible to increase the odds that we reach close to the real optimal value.

After obtaining $N_\mr{GHZ}^*$ (at $\eta = \eta_0$), we consider the logical identity gate with $T = d - 1$ (see Supplementary Figure~\ref{fig:identity_gate}) to calculate $\mathcal{N}_{p_L^\mr{targ}}$.
$\mathcal{N}_{p_L^\mr{targ}}$ is determined by the following equality:
\begin{align*}
    \mathcal{N}_{p_L^\mr{targ}} = N_\mr{GHZ}^* (2d_{p_L^\mr{targ}} + 1)(3d_{p_L^\mr{targ}}^2 - 3d_{p_L^\mr{targ}} + 1),
\end{align*}
where $d_{p_L^\mr{targ}}$ is the minimal code distance to achieve the target logical error rate of $p_L^\mr{targ}$ for the identity gate when $\eta = \eta_0$.
$d_{p_L^\mr{targ}}$ is obtained by employing the error simulation method in Supplementary Note~2.
However, this method simulates the logical identity gate with $T=4d+1$, while our current interest is that with $T = d - 1$.
From an obtained logical error rate $p_L'$ from the method with $T=4d+1$, we estimate the \textit{logical error rate} $p_L^{(1)}$ \textit{per two layers (one unit cell)} from the relation
\begin{align*}
    p_L' = \sum_{t \leq T: \mr{odd}} \binom{T}{t} \qty(p_L^{(1)})^t \qty(1 - p_L^{(1)})^{T - t} = \frac{1}{2} \qty[ 1 - \qty(1 - 2p_L^{(1)})^{T} ],
\end{align*}
where $T = 4d+1$.
Using a similar relation for $T = d-1$, we can convert $p_L'$ into the logical error rate $p_L$ of the gate with $T = d-1$.
We then obtain $d_{p_L^\mr{targ}}$ by calculating the logical error rate $p_L^{(d)}$ at $\eta = \eta_0$ for each code distance $d \leq 11$ and finding the smallest $d$ satisfying $p_L^{(d)} < p_L^\mr{targ}$.
If $p_L^{(11)} \geq p_L^\mr{targ}$, $d_{p_L^\mr{targ}}$ is estimated from the linear extrapolation of the points $(9,~\log p_L^{(9)})$ and $(11,~\log p_L^{(11)})$.

\newpage

\section*{Supplementary Note~4: Analysis of the approach using single-photon qubits with fusions assisted by ancillary photons}
\label{SN:analysis_single_photon}

We here investigate the approach using single photons for all qubits with fusions assisted by ancillary photons \cite{ewert2014efficient}.
We consider an RHG lattice generated by the process shown in the main text, where every fusion is done by the scheme in Ref.~\cite{ewert2014efficient}.
We do not use the purification process in Ref.~\cite{herr2018local}, which is worth investigating in future works.
Under the noise model described in the main text, a fusion detects a loss with probability $1 - (1 - \eta)^2$, if losses in ancillary photons are neglected.
Since a marginal state of every Bell state is maximally mixed, detection of a photon loss means complete loss of information; thus, $q_\mr{lett} = q_\mr{sign} = 1/2$ in such a case.
If losses are not detected, the fusion fails with probability $p_\mr{f}$, where the letter information of the Bell state still can be obtained \cite{ewert2014efficient}; namely, $q_\mr{lett} = 0$ and $q_\mr{sign} = 1/2$.
These two cases make some central qubits deficient, which can be tracked using the methodology of analyzing nonideal fusions that is presented in the main text.
HIC is used for the $H$-configuration to make the failure of a step-1 fusion affects only one central qubit; see Fig.~3 of the main text.

\begin{figure}[ht!]
    \centering
    \includegraphics[width=0.47961631\textwidth]{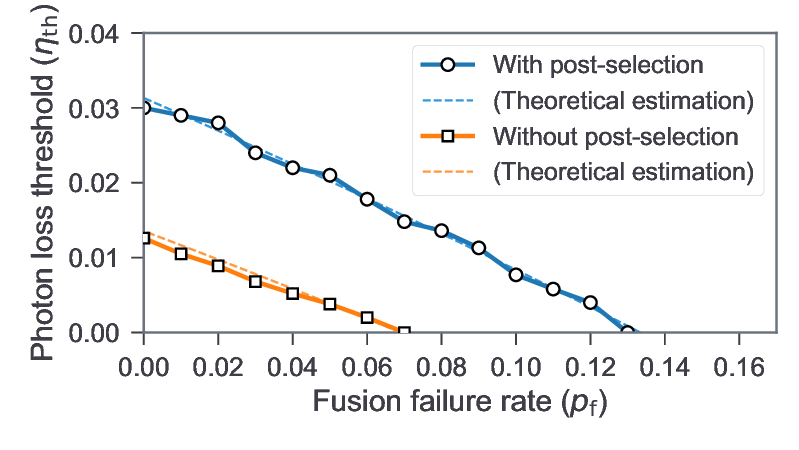}
    \caption{\textbf{Simulation results for the approach using single-photon qubits with fusions assisted by ancillary photons.}
    It shows the photon loss thresholds $\eta_\mr{th}$ obtained from simulations or estimated theoretically as a function of the fusion failure rate $p_\mr{f}$.}
    \label{fig:threshold_no_encoding}
\end{figure}

The photon loss thresholds calculated numerically are plotted in Supplementary Figure~\ref{fig:threshold_no_encoding} with theoretical estimations for various values of $p_\mr{f}$.
The theoretical estimation is done by the following methods:
We first assume that star clusters are not post-selected.
For a central qubit~$q$ to be not deficient, the following conditions should be satisfied simultaneously:
\begin{enumerate}
    \item Two step-1 fusions in the star cluster containing $q$ succeed. 
    \item Four step-1 fusions in the four adjacent star clusters (one for each) do not detect losses.
    \item Four step-2 fusions involved in the star cluster containing $q$ do not detect losses. Two among them (that make $q$ deficient if they fail) succeed. 
    \item $q$ itself does not suffer a loss.
\end{enumerate}
From above, we obtain the probability that a central qubit in the final lattice is intact: $p_\mr{int}\qty(\eta, p_\mr{f}) = \qty( 1 - p_\mr{f} )^4 (1 - \eta)^{21}$.
If star clusters are post-selected, the first and second conditions are no longer needed, thus we get $p_\mr{int}\qty(\eta, p_\mr{f}) = \qty( 1 - p_\mr{f} )^2 (1 - \eta)^9$.
Regarding a 50\% chance of a $Z$-error as erasing the qubit by measuring it in the $Z$-basis (while ignoring the correlation of errors), a photon loss threshold $\eta_\mr{th}$ can be estimated by solving $1 - p_\mr{prc} = p_\mr{int}\qty(\eta_\mr{th}, p_\mr{f})$, where $p_\mr{prc} = 0.249$ is the known cubic-lattice bond percolation threshold \cite{barrett2010fault, lorenz1998precise}.

Supplementary Figure~\ref{fig:threshold_no_encoding} shows that $p_\mr{f}$ should be less than about 10\% (1\%) even if $\eta$ is only 1\% when star clusters are (are not) post-selected.
The failure rate of $10\%$ can be achieved by using the BSM scheme of $N=3$ in Ref.~\cite{ewert2014efficient} where $p_\mr{f} = 6.25\%$.
For one BSM with $N=3$, 32 photon-number resolving detectors (PNRDs) resolving up to 16 photons and the ancillary states $\ket{\Upsilon_1}, \ket{\Upsilon_2}, \ket{\Upsilon_3}$ (two copies each) are required.
$\ket{\Upsilon_j}$ is a $2^j$-mode state defined as $\ket{\Upsilon_j} := \ket{2, 0, 2, 0, \cdots, 2, 0} + \ket{0, 2, 0, 2, \cdots, 0, 2}$.
Although $\ket{\Upsilon_1}$ can be simply generated with two identical single photons and a beam splitter using Hong-Ou-Mandel effect \cite{hong1987measurement}, it is probably impossible to obtain $\ket{\Upsilon_j}$'s for $j \geq 2$ from single photons with linear optics \cite{ewert2014efficient}.
Therefore, it is highly demanding to implement this MBQC protocol with linear optics due to the requirements of PNRDs that can resolve many photons and ancillary states hard to generate.

\newpage

\section*{Supplementary Note~5: Analysis of the approach using simple repetition codes}
\label{SN:analysis_repetition_codes}

We here investigate the approach using simple repetition codes, which is covered in our previous work \cite{omkar2022all}.
In this protocol (called ``MTQC''), side qubits are $n$-photon ones encoded in the basis of $\qty{ \ket{\textsc{h}}^{\otimes n}, \ket{\textsc{v}}^{\otimes n}}$, where $n$ is a natural number.
For central qubits, we first consider using $m$-photon qubits and then concatenate them with the $N$-repetition code.
That is, we use the basis of $\qty{\qty(\ket{\textsc{h}}^{\otimes m} \pm \ket{\textsc{v}}^{\otimes m})^{\otimes N}}$ for the central qubits.
In Ref.~\cite{omkar2022all}, the photon loss thresholds and resource overheads are analyzed in detail, but a rigorous analysis of the effects of nonideal fusions like that done for PTQC is lacking.

Since the $n$-photon encoding for side qubits is equal to the $(n, 1)$ parity encoding, the effects of nonideal fusions can be analyzed in the same way as done for PTQC with the $(n, 1)$ parity encoding.
The difference between the two is the way that central qubits become deficient due to photon losses in themselves.
In PTQC, central qubits are single photons, thus a central qubit becomes deficient with probability $\eta$.
In MTQC, however, the deficiency rate due to photon losses in central qubits is $\qty[1 - (1 - \eta)^m]^N$, which decreases exponentially as $N$ increases.
The photon loss thresholds recalculated based on these facts are presented in Supplementary Figure~\ref{fig:mtqc_comparison} with the previous values reported in Ref.~\cite{omkar2022all}, which shows that the recalculated photon loss thresholds are smaller than the reported values.
In particular, it is observed that the central qubit encoding strategy does not improve the thresholds significantly.
This discrepancy is because the detrimental effects of nonideal fusion affecting nearby qubits have not been sufficiently rigorously addressed.

\begin{figure}[ht!]
    \centering
    \includegraphics[width=0.95\textwidth]{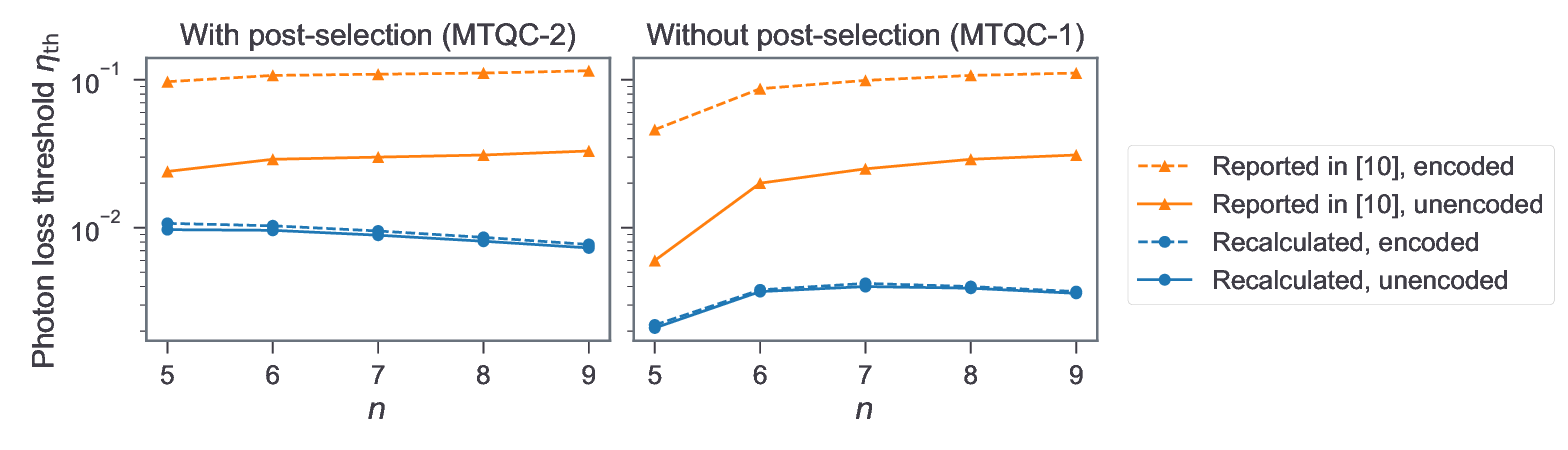}
    \caption{
        \textbf{Simulation results for the approach using the simple repetition codes.}
        It shows the photon loss thresholds $\eta_\mr{th}$ as a function of $n$ for MTQC, which are obtained from Ref.~\cite{omkar2022all} and the recalculation using the methodology for analyzing nonideal fusions.
        Other parameters are $(m, N) = (2, 1)$ and $(m, N) = (2, 3)$ for the unencoded and encoded cases, respectively.
        Two subvariants of MTQC, one with the post-selection of star clusters and the other without it, are considered, which are respectively termed MTQC-2 and MTQC-1 in Ref.~\cite{omkar2022all}.
    }
    \label{fig:mtqc_comparison}
\end{figure}

\newpage

\section*{Supplementary Note~6: Calculation of the error probabilities of a CBSM when on-off detectors are used}
\label{SN:cbsm_error_probabilities}

We here derive the error probabilities of a CBSM on two qubits (say, qubits 1 and 2) encoded with the parity encoding when on-off detectors are used for fusions.
We denote $x := (1 - \eta)^2$, which is the probability that a \bsmp~does not detect photon losses.
Note the following decomposition of Bell states:
\begin{align}
    \ket{\Phi^\pm} &= 2^{-\frac{n-1}{2}} \sum_{l:\mathrm{even(odd)} \leq n} \mathcal{P} \qty[ \ket{\phi_{(m)}^-}^{\otimes l} \ket{\phi_{(m)}^+}^{\otimes n-l} ], \label{eq:logical_phi_decomposition} \\
    \ket{\Psi^\pm} &= 2^{-\frac{n-1}{2}}\sum_{l:\mathrm{even(odd)} \leq n} \mathcal{P} \qty[ \ket{\psi_{(m)}^-}^{\otimes l} \ket{\psi_{(m)}^+}^{\otimes n-l} ], \label{eq:logical_psi_decomposition} \\
    \ket{\phi_{(m)}^\pm} &= 2^{-\frac{m-1}{2}} \sum_{k:\mathrm{even} \leq m} \mathcal{P} \qty[ \ket{\psi^\pm}^{\otimes k} \ket{\phi^\pm}^{\otimes m-k} ], \label{eq:block_phi_decomposition} \\
    \ket{\psi_{(m)}^\pm} &= 2^{-\frac{m-1}{2}} \sum_{k:\mathrm{odd} \leq m} \mathcal{P} \qty[ \ket{\psi^\pm}^{\otimes k} \ket{\phi^\pm}^{\otimes m-k} ], \label{eq:block_psi_decomposition}
\end{align}

\subsection{Block-level BSM (\bsmb)}

We note that every positive operator-valued measure (POVM) element of a lossy \bsmp~has vanishing off-diagonal entries in the Bell basis; see the last subsection of this Supplementary Note for the proof.
Also, each POVM element of a lossy \bsmb, denoted by $M^\mr{blc}_O$ for each outcome $O = (r, s, \vb{U})$, is the tensor product of particular POVM elements of the lossy \bsmp's constituting the \bsmb.
Thus, the conditional probability of getting $O$ from a block-level Bell state $\ket{B}$ is
\begin{align*}
    \Pr\qty(O  \middle|  B) = \bra{B} M^\mr{blc}_O \ket{B} = \frac{1}{2^{m-1}} \sum_i \bra{B_i} M^\mr{blc}_O \ket{B_i} = \frac{1}{2^{m-1}} \sum_i \Pr\qty(O \middle| B_i),
\end{align*}
where $\ket{B_i}$'s are the terms constituting the summation in Eqs.~\eqref{eq:block_phi_decomposition} and \eqref{eq:block_psi_decomposition}, namely, $\ket{B} = \frac{1}{\sqrt{2^{m-1}}} \sum_i \ket{B_i}$.
In other words, when calculating $\Pr\qty(O  \middle|  B)$, it is enough to find $\Pr\qty(O \middle| B_i)$'s and then take their average.

The posterior probability of a block-level Bell state $\ket{B}$ under a given outcome $O$ is
\begin{align}
    \Pr\qty(B  \middle|  O) = \frac{\Pr\qty(O  \middle|  B)}{\sum_{\ket{B'} \in \mathcal{B}_\mr{blc}} \Pr\qty(O \middle| B')}, \label{supeq:posterior_prob}
\end{align}
where $\mathcal{B}_\mr{blc}$ is the set of the four block-level Bell states.
Thus, the result of the \bsmb~is selected randomly in the set $R(O) := \mr{argmax}_B \Pr\qty(O \middle| B)$.
The sign (letter) error probability as a function of $O$ is
\begin{align}
\begin{split}
    q^\mr{blc}_\mr{sign(lett)} (O) = \frac{1}{\abs{R(O)}} \sum_{\ket{B} \in R(O)} \qty[ \Pr\qty(F_\mr{sign(lett)} (B) \middle| O) + \Pr\qty(F_\mr{sign} \circ F_\mr{lett} (B) \middle| O)],     
\end{split}
\label{supeq:error_prob_general}
\end{align}
where $\ket{F_\mr{sign(lett)} (B)}$ is the Bell state obtained by flipping the sign (letter) from $\ket{B}$ (e.g., $F_\mr{sign} (\phi^\pm) = \phi^\mp$).

Block-level outcomes can be grouped by the $j+3$ events $\mathcal{S}_r~(r=0, \cdots, j)$, $\mathcal{F}$, and $\mathcal{D}$, as defined in the main text.
We now calculate the probability that each event occurs and the corresponding sign and letter error probabilities.
Let us first consider an outcome $O = (r, s, \vb{U}) \in \mathcal{S}_r$, where $N_\mr{f} (\vb{U}) = 0$.
Regarding a single term in the decomposition of $\ket{\phi^\pm_{(m)}}$ [see Eq.~\eqref{eq:block_phi_decomposition}], if there are total $k$ of $\ket{\psi^\pm}$'s, the first $r$ physical levels contain $k - N_\psi(\vb{U})$ of $\ket{\psi^\pm}$'s, which should suffer photon losses by the definition of $r$.
If $r < j$, $s$ selected by the successful $(r+1)$th $B_\psi$ is certainly $\pm$, the sign of $\ket{\phi^\pm_{(m)}}$.
If $r = j$, the randomly selected $s$ may or may not be corrected; however, the latter case is out of $\mathcal{S}_r$ since all the following $B_\mp$'s must fail.
The remaining $m-r$ \bsmp's should not suffer photon losses since $N_f(\vb{U}) = 0$.
Hence, for all $\vb{U}$ satisfying $N_\mr{f} (\vb{U}) = 0$, we get
\begin{equation}
\begin{split}
    \Pr\qty(r, \pm, \vb{U}  \middle|  \phi^\pm_{(m)}) &= \frac{1}{2^{m-1}} \sum_{k:\mr{even} \leq r + N_\psi} \binom{r}{k - N_\psi} \qty(1 - x)^{k - N_\psi} \frac{1}{2^{\delta_{rj}}} x^{m - r} \\
    &= \frac{1}{2^{\delta_{rj}}} \qty[\qty(1 - \frac{x}{2})^r \qty(\frac{x}{2})^{m-r} + (-1)^{N_\psi} \qty(\frac{x}{2})^m], \\
    \Pr\qty(r, \mp, \vb{U}  \middle|  \phi^\pm_{(m)}) &= 0,
\end{split}
\label{supeq:indv_prob_Ar_phi}
\end{equation}
where $N_\psi = N_\psi(\vb{U})$.
Similarly, we get
\begin{equation}
\begin{split}
    \Pr\qty(r, \pm, \vb{U} \middle| \psi^\pm_{(m)}) &= \frac{1}{2^{\delta_{rj}}} \qty[\qty(1 - \frac{x}{2})^r \qty(\frac{x}{2})^{m-r} - (-1)^{N_\psi} \qty(\frac{x}{2})^m], \\
    \Pr\qty(r, \mp, \vb{U} \middle| \psi^\pm_{(m)}) &= 0.
\end{split}
\label{supeq:indv_prob_Ar_psi}
\end{equation}
From Eqs.~\eqref{supeq:posterior_prob}--\eqref{supeq:indv_prob_Ar_psi}, we obtain
\begin{equation}
\begin{split}
    q^\mr{blc}_\mr{sign} \qty(O) &= 0 =: q^\mr{blc}_\mr{sign}\qty(\mathcal{S}_r), \\
    q^\mr{blc}_\mr{lett} \qty(O) &= \frac{ \qty(1 - \frac{x}{2})^r \qty(\frac{x}{2})^{m-r} - \qty(\frac{x}{2})^m }{ 2 \qty(1 - \frac{x}{2})^r \qty(\frac{x}{2})^{m-r} } = \frac{1}{2} \qty[ 1 - \qty(\frac{x}{2 - x})^r ] =: q^\mr{blc}_\mr{lett}\qty(\mathcal{S}_r).
\end{split}
\label{supeq:error_probs_Ar}
\end{equation}
Note that the error probabilities are the same for all $O \in \mathcal{S}_r$.
The total probability that the event $\mathcal{S}_r$ occurs is
\begin{equation*}
    p_{\mathcal{S}_r} := \frac{1}{4} \sum_{O \in \mathcal{S}_r} \sum_{\ket{B} \in \mathcal{B}_\mr{blc}} \Pr\qty(O  \middle|  B) = \frac{1}{2} \frac{1}{2^{\delta_{rj}}} \qty(1 - \frac{x}{2})^r \qty(\frac{x}{2})^{m-r}  2^{m - r - 1 + \delta_{rj}} = \frac{1}{2} \qty(1 - \frac{x}{2})^r x^{m-r} ,
\end{equation*}
where the factor $2^{m - r - 1 + \delta_{rj}}$ is the number of possible $\vb{U}$'s for a given value of $r$.

Next, we consider $O = (j, s, \vb{U}) \in \mathcal{F}$, where $N_f(\vb{U}) = m-j$, namely, all the $B_s$'s fail.
Regarding a single term in the decomposition of $\ket{\phi^\pm_{(m)}}$, all the $\ket{\psi^\pm}$'s in the first $j$ physical levels should suffer photon losses.
If $s = \pm$, all the following $B_\pm$'s should suffer photon losses as well.
If $s = \mp$, all the following $B_\mp$'s fail regardless of photon losses.
We thus get
\begin{equation*}
\begin{split}
    \Pr\qty(j, \pm, \vb{U} = (f, \cdots, f) \middle| \phi^\pm_{(m)}) &= \frac{1}{2^{m-1}} \sum_{\substack{0 \leq k_1 \leq j \\ 0 \leq k_2 \leq m-j \\ k_1 + k_2:~\mr{even}}} \binom{j}{k_1} \binom{m-j}{k_2} (1 - x)^{k_1 + m - j} \cdot \frac{1}{2} = \frac{1}{2} \qty(1 - \frac{x}{2})^j (1 - x)^{m-j}, \\
    \Pr\qty(j, \mp, \vb{U} = (f, \cdots, f) \middle| \phi^\pm_{(m)}) &= \frac{1}{2} \qty(1 - \frac{x}{2})^j,
\end{split}
\end{equation*}
where $k_1$ ($k_2$) in the summation indicates the number of $\ket{\psi^\pm}$'s in the first $j$ (last $m-j$) physical levels.
Similarly, the same results are obtained for $\ket{\psi^\pm_{(m)}}$:
\begin{equation*}
\begin{split}
    \Pr\qty(j, \pm, \vb{U} = (f, \cdots, f) \middle| \psi^\pm_{(m)}) &= \frac{1}{2} \qty(1 - \frac{x}{2})^j (1 - x)^{m-j}, \\
    \Pr\qty(j, \mp, \vb{U} = (f, \cdots, f) \middle| \psi^\pm_{(m)}) &= \frac{1}{2} \qty(1 - \frac{x}{2})^j.
\end{split}
\end{equation*}
The corresponding error probabilities are
\begin{equation*}
\begin{split}
    q^\mr{blc}_\mr{sign}(O) = \frac{ \qty( 1 - x )^{m - j} }{ 1 + \qty( 1 - x )^{m-j}} =: q^\mr{blc}_\mr{sign}(\mathcal{F}), \qquad q^\mr{blc}_\mr{lett}(O) = \frac{1}{2} =: q^\mr{blc}_\mr{lett}(\mathcal{F})
\end{split}
\end{equation*}
and the total probability of the event $\mathcal{F}$ is
\begin{align*}
    p_{\mathcal{F}} = \frac{1}{4} \sum_{s = \pm} \sum_{\ket{B} \in \mathcal{B}_\mr{blc}} \Pr\qty(j, s, \vb{U} = (f, \cdots, f) \middle|  B) = \frac{1}{2}\qty(1 - \frac{x}{2})^j \qty[ 1 + (1 - x)^{m-j} ].
\end{align*}

Lastly, we consider $O = (r, s, \vb{U}) \in \mathcal{D}$.
If $r < j$, $N_f(\vb{U}) > 0$ by the definition of $\mathcal{D}$ and $N_f(\vb{U}) < m - r$ since the first component of $\vb{U}$ is always $\psi$.
If $r = j$, $0 < N_f(\vb{U}) < m - j$ by the definition of $\mathcal{D}$.
Therefore, regardless of $r$, $\vb{U}$ contains at least one failure and one success ($\psi$ or $\phi$).
Thanks to the successful \bsmp's, the sign of the result is identified without an error.
On the other hand, the letter is not identified because of the failures.
We can see intuitively without calculation that the letter error probability is $1/2$:
Even if there is only one failure in $\vb{U}$, the letter information of the corresponding physical-level Bell state is completely lost, considering that the marginal state of a block-level Bell state on a single physical level is $\ketbra{\phi^\pm} + \ketbra{\psi^\pm}$.
Thus, the block-level letter information (determined by the parity of the number of \bsmp~outcomes with $\psi$) is completely lost as well.
To rewrite the results, we get
\begin{align*}
    q^\mr{blc}_\mr{sign}(\mathcal{D}) = 0, \qquad q^\mr{blc}_\mr{lett}(\mathcal{D}) = \frac{1}{2}, \qquad p_\mathcal{D} = 1 - \sum_{r=0}^j p_{\mathcal{S}_r} - p_\mathcal{F}.
\end{align*}

\subsection{Lattice-level BSM (\bsml)}

Each $n$-tuple of events composed of $\mathcal{S}_r$ ($0 \leq r \leq j$), $\mathcal{F}$, and $\mathcal{D}$ corresponds to a set of possible outcomes of a \bsml.
Let us consider such an $n$-tuple $\mathbf{E} = \qty(\mathcal{E}_1, \cdots, \mathcal{E}_n)$.
A lattice-level sign error occurs when there is an odd number of block-level sign errors and $\mathcal{F}$ is the only event where a block-level sign error may occur; thus, the sign error probability is
\begin{align*}
    q_\mr{sign} = \sum_{i:\mathrm{odd} \leq N_\mathcal{F}} \binom{N_\mathcal{F}}{i} q_\mr{sign}^\mr{blc}(\mathcal{F})^i \qty[1 - q_\mr{sign}^\mr{blc}(\mathcal{F})]^{N_\mathcal{F} - i} = \frac{1}{2} - \frac{1}{2} \qty[1 - 2q_\mr{sign}^\mr{blc}(\mathcal{F})]^{N_\mathcal{F}},
\end{align*}
where $N_\mathcal{F}$ is the number of $\mathcal{F}$'s in $\mathbf{E}$.

A lattice-level letter error occurs when the weighted majority vote of the block-level letters gives a wrong answer.
We consider i.i.d. random variables $\Lambda_1, \cdots, \Lambda_n$ such that $\Lambda_i \sim \mr{Bernoulli}\qty(q_i)$ for each $i$ where $q_i := q^\mr{blc}_\mr{lett}\qty(\mathcal{E}_i)$, which indicates whether a letter error occurs in the $i$th block.
A lattice-level letter error occurs if
\begin{align*}
    \sum_i (2 \Lambda_i - 1) \log \frac{1-q_i}{q_i} =: V(\Lambda_1, \cdots, \Lambda_n)
\end{align*}
is larger than zero or if it is equal to zero and the randomly selected letter is wrong.
Therefore, we get
\begin{align*}
    q_\mr{lett} &= \Pr\qty(V(\Lambda_1, \cdots, \Lambda_n) > 0) + \frac{1}{2} \Pr\qty(V(\Lambda_1, \cdots, \Lambda_n) = 0) \\
    &= \sum_{\qty(\lambda_1, \cdots, \lambda_n) \in \mathbb{Z}_2^n} \prod_{i=1}^n \Pr\qty(\Lambda_i = \lambda_i) \qty{ \Theta\qty[V\qty(\lambda_1, \cdots, \lambda_n) > 0] + \frac{1}{2} \Theta\qty[V\qty(\lambda_1, \cdots, \lambda_n) = 0] } \\
    &= \sum_{(\lambda_1, \cdots, \lambda_n) \in \mathbb{Z}_2^n} \prod_{i=1}^n \qty[q_i^{\lambda_i} \qty(1 - q_i)^{1 - \lambda_i}] \qty[ \frac{1}{2} \mr{sgn}\qty( V\qty(\lambda_1, \cdots, \lambda_n) ) + \frac{1}{2}], \\
    &= \frac{1}{2} + \frac{1}{2} \sum_{(\lambda_1, \cdots, \lambda_n) \in \mathbb{Z}_2^n} \prod_{i=1}^n \qty[ q_i^{\lambda_i} \qty(1 - q_i)^{1 - \lambda_i} ] \mr{sgn}\qty( \sum_{i=1}^n \qty(2\lambda_i - 1) \log\frac{1 - q_i}{q_i} ),
\end{align*}
where $\Theta[C]$ for a condition $C$ is equal to 1 if $C$ is true and 0 if it is false, and $\mr{sgn}(a)$ is $a/\abs{a}$ if $a \neq 0$ and 0 if $a = 0$.

\subsection{Proof of vanishing off-diagonal entries of the POVM elements of a lossy \bsmp}
\label{subsec:vanishing_off_diagonal_entries}

We here prove that every POVM element of a lossy \bsmp~in a CBSM with on-off detectors has vanishing off-diagonal entries in the Bell basis.
Let $\Lambda_\eta$ denote the photon loss channel of a loss rate $\eta$ defined as $\Lambda_\eta (\sigma) := (1 - \eta) \sigma + \eta \ketbra{0}$ for a single-photon state $\sigma$ and the vacuum state $\ket{0}$.
By substituting $\sigma = \ketbra{\psi}$ for an arbitrary pure state $\ket{\psi} = \alpha \ket{\textsc{h}} + \beta \ket{\textsc{v}}$, we get $\Lambda_\eta (\ketbra{H}{V}) = (1 - \eta) \ketbra{H}{V}$. Thus,
\begin{equation}
\begin{split}
    (\Lambda_\eta \otimes \Lambda_\eta)  (\ketbra{\phi^\pm}{\psi^\pm}) &= (1 - \eta)^2 \ketbra{\phi^\pm}{\psi^\pm} + \eta (1 - \eta) \qty(\ketbra{H0}{V0} + \ketbra{V0}{H0} + \ketbra{0H}{0V} + \ketbra{0V}{0H}), \\
    (\Lambda_\eta \otimes \Lambda_\eta)  (\ketbra{\phi^\pm}{\psi^\mp}) &= (1 - \eta)^2 \ketbra{\phi^\pm}{\psi^\mp} + \eta (1 - \eta) \qty(\ketbra{H0}{V0} - \ketbra{V0}{H0} - \ketbra{0H}{0V} + \ketbra{0V}{0H}), \\
    (\Lambda_\eta \otimes \Lambda_\eta)  (\ketbra{\phi^+}{\phi^-}) &= (1 - \eta)^2 \ketbra{\phi^+}{\phi^-} + \eta (1 - \eta) (\ketbra{H0} - \ketbra{V0} + \ketbra{0H} - \ketbra{0V}), \\
    (\Lambda_\eta \otimes \Lambda_\eta)  (\ketbra{\psi^+}{\psi^-}) &= (1 - \eta)^2 \ketbra{\psi^+}{\psi^-} + \eta (1 - \eta) (\ketbra{H0} - \ketbra{V0} - \ketbra{0H} + \ketbra{0V}).
\end{split}
\label{supeq:bell_off_diagonal_photon_loss}
\end{equation}
Now, let $M_\pm$ and $M_f$ denote the POVM elements of a lossy $B_\psi$ corresponding to the outcomes $\ket{\psi^\pm}$ and failure, respectively.
By modelling a lossy $B_\psi$ as a photon loss channel followed by an ideal $B_\psi$, $M_i$ for $i \in \qty{+, -, f}$ satisfies
\begin{align}
    \Tr[M_i \rho] = \Tr[\Pi_i (\Lambda_\eta \otimes \Lambda_\eta) (\rho)],
    \label{supeq:povm_lossy_bsm}
\end{align}
for any two-qubit state $\rho$, where
\begin{align*}
    \Pi_+ :=& \ketbra{\psi^+}, \qquad \Pi_- := \ketbra{\psi^-}, \\
    \Pi_f :=& \ketbra{\phi_+} + \ketbra{\phi_-} + \ketbra{H0} + \ketbra{V0} + \ketbra{0H} + \ketbra{0V} + \ketbra{00}
\end{align*}
are the projectors of an ideal $B_\psi$ with a lossy input.
From Eqs.~\eqref{supeq:bell_off_diagonal_photon_loss} and \eqref{supeq:povm_lossy_bsm}, we obtain $\bra{\psi^\pm} M_i \ket{\phi^\pm} = \bra{\psi^\pm} M_i \ket{\phi^\mp} = \bra{\phi^+} M_i \ket{\phi^-} = \bra{\psi^+} M_i \ket{\psi^-} = 0$ for every $i \in \qty{+, -, f}$.
Similar arguments can be done for a lossy $B_+$ and $B_-$ as well.

\newpage

\section*{Supplementary Note~7: Derivation of the physical-level graphs of post-$H$ microclusters}
\label{SN:physical_level_graph}

\begin{figure}[ht!]
    \centering
    \includegraphics[width=\columnwidth]{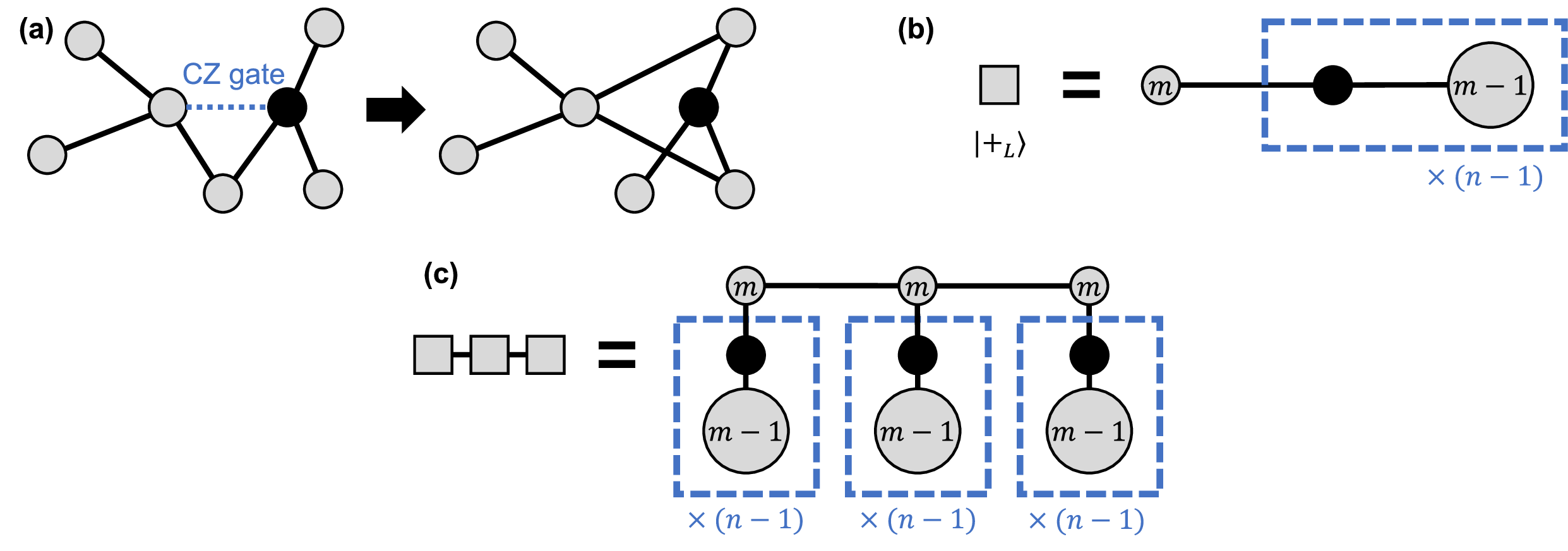}
    \caption{
    \textbf{Different steps of deriving the physical-level graphs of post-$H$ microclusters.}
    (a) Transformation of a graph state by applying a Hadamard gate followed by applying a \cz~gate.
    (b) Physical-level graph structure of the state $\ket{+_L} = \ket{0_L} + \ket{1_L}$.
    (c) Physical-level graph structure of a lattice-level three-qubit linear graph state.
    }
    \label{fig:graph_structure_steps}
\end{figure}

In this note, we derive the physical-level graph structures of the central and side post-$H$ microclusters for the two $H$-configurations, which are shown in Fig.~7 of the main text.
The first step of the derivation is to investigate how a graph state is transformed if a Hadamard gate ($H_1$) is applied on one of the qubits (say, qubit~1) and then a \cz~gate ($C^Z_{12}$) is applied on qubit~1 and another qubit (say, qubit~2) that is not adjacent to qubit~1.
Note that, in the Heisenberg picture, the \cz~gate transforms the Pauli-$X$ operators of the qubits as $X_1 \rightarrow X_1 Z_2$ and $X_2 \rightarrow Z_1 X_2$, while it leaves the Pauli-$Z$ operators the same.
For a qubit~$i$, $S_i := X_i \prod_{j \in N(i)} Z_{j}$, where $N(i)$ is the set of qubits adjacent to qubit~$i$, is a stabilizer of the initial graph state.
The stabilizers $S_1$ and $S_1 S_2$ are transformed by $C^Z_{12}H_1$ as
\begin{align*}
    S_1 = X_1 \prod_{j \in N(1)} Z_j &\longrightarrow Z_1 \prod_{j \in N(1)} Z_j = H_1 \qty( X_1 \prod_{j \in N(1)} Z_j ) H_1, \\
    S_1 S_2 = X_1 X_2 \prod_{j \in N(1) \triangle N(2)} Z_j &\longrightarrow X_2 \prod_{j \in N(1) \triangle N(2)} Z_j = H_1 \qty( X_2 \prod_{j \in N(1) \triangle N(2)} Z_j ) H_1,
\end{align*}
where $A \triangle B := A \cup B \setminus \qty( A \cap B )$ for two sets $A$ and $B$.
Also, for each qubit~$i \in N(1)$,
\begin{align*}
    S_i = X_i \prod_{j \in N(i)} Z_j \longrightarrow X_i X_1 Z_2 \prod_{j \in N(i) \setminus \qty{1}} Z_j = H_1 \qty( X_i \prod_{j \in N(i) \triangle \qty{2}} Z_j ) H_1.
\end{align*}
Therefore, the overall effect of the process is, for each qubit~$i$ adjacent to qubit~1, to flip the connectivity of the qubits 2 and $i$ (namely, connect them if they are disconnected and disconnect them if they are already connected) and then apply $H_1$.
An example of this transformation is presented in Supplementary Figure~\ref{fig:graph_structure_steps}(a).

\begin{figure}[ht!]
    \centering
    \includegraphics[width=0.44726302\columnwidth]{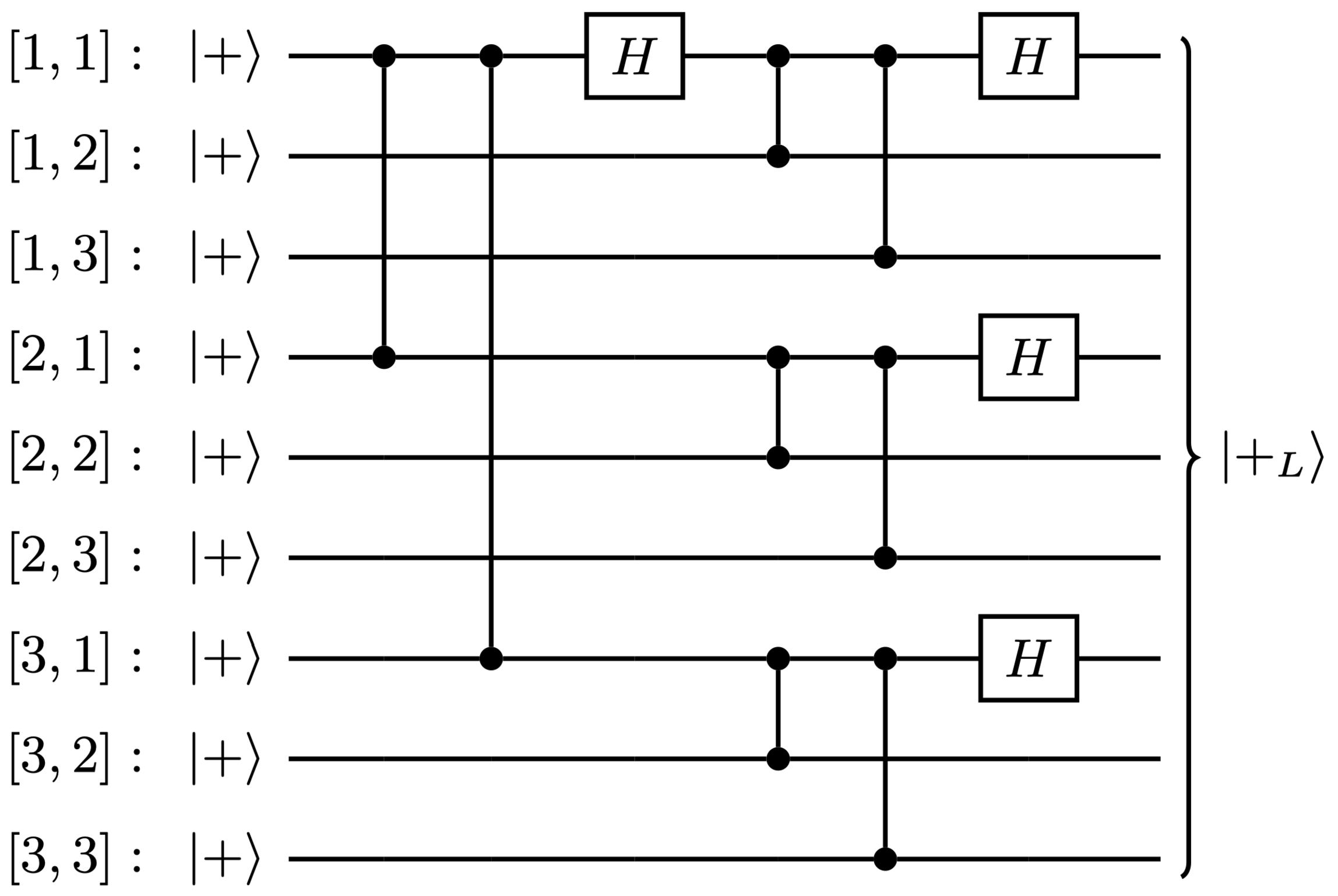}
    \caption{
        \textbf{Encoding circuit of the state $\ket{+_L} := \ket{0_L} + \ket{1_L}$ in the $(3, 3)$ parity encoding.}
        It employs multiple copies of the state $\ket{+} := \ket{\textsc{h}} + \ket{\textsc{v}}$, \cz~gates, and Hadamard gates.
        The label $[i, j]$ for each physical-level qubit indicates the index $i$ of the block and the index $j$ of the photon in the block.
    }
    \label{fig:encoding_circuit}
\end{figure}

Next, we obtain the graph structure of the state $\ket{+_L} := \ket{0_L} + \ket{1_L}$.
Supplementary Figure~\ref{fig:encoding_circuit} shows the encoding circuit of the state for the $(3, 3)$ parity encoding, which employs multiple copies of the state $\ket{+} := \ket{\textsc{h}} + \ket{\textsc{v}}$, \cz~gates, and Hadamard gates.
Here, we label the $j$th physical qubit of the $i$th block by $[i, j]$.
It is straightforward to generalize it for any pair of $(n, m)$
The graph structure of $\ket{+_L}$ shown in Supplementary Figure~\ref{fig:graph_structure_steps}(b) is obtained by preparing $nm$ isolated vertices and tracking the transformation of the graph via the \cz~and Hadamard gates in the circuit.

A lattice-level \cz~gate $C^Z_L$ is done by $m^2$ physical \cz~gates:
\begin{align}
    C^Z_L = \prod_{i, j \leq m} C^Z_{1i, 1j},
    \label{supeq:logical_cz_gate}
\end{align}
where $C^Z_{ij, kl}$ is the \cz~gate between the $[i, j]$ qubit of the first lattice-level qubit and the $[k, l]$ qubit of the second lattice-level qubit.
It can be verified as follows:
The stabilizer generators of the $(n, m)$ parity encoding are
\begin{align*}
    \qty{X_{ij} X_{i(j+1)}~~(\forall i \leq n,~ \forall j \leq m - 1), \quad \prod_{j=1}^m Z_{ij} Z_{(i+1)j}~~(\forall i \leq n-1)}
\end{align*}
and the lattice-level Pauli operators are
\begin{align*}
    X_L = X_{11} \cdots X_{n1}, \qquad Z_L = Z_{11} \cdots Z_{1m},
\end{align*}
where $X_{ij}$ ($Z_{ij}$) is the Pauli-$X$ (-$Z$) operator on the $[i, j]$ qubit.
It is straightforward to see that the RHS of Eq.~\eqref{supeq:logical_cz_gate} commutes with all the stabilizers and transforms the lattice-level Pauli operators correctly.

Combining the above results on the $\ket{+_L}$ state and the lattice-level \cz~gate, we attain the graph structure of a lattice-level three-qubit linear graph state shown in Supplementary Figure~\ref{fig:graph_structure_steps}(c).
The only left ingredient is the lattice-level Hadamard gate ($H_L$).
The circuit for $H_L$ is obtained by simply connecting the decoding circuit, the physical Hadamard gate, and the encoding circuit, which is explicitly shown in Supplementary Figure~\ref{fig:logical_hadamard_circuit} for the (3, 3) parity encoding.
By transforming the graph in Supplementary Figure~\ref{fig:graph_structure_steps}(c) with appropriate lattice-level Hadamard gates, we finally get the desired graph structures of the post-$H$ microclusters shown in Fig.~7 of the main text.
Note that, for central microclusters, the middle lattice-level qubits are replaced with unencoded physical-level qubits.

\begin{figure}[ht!]
    \centering
    \includegraphics[width=0.62216288\textwidth]{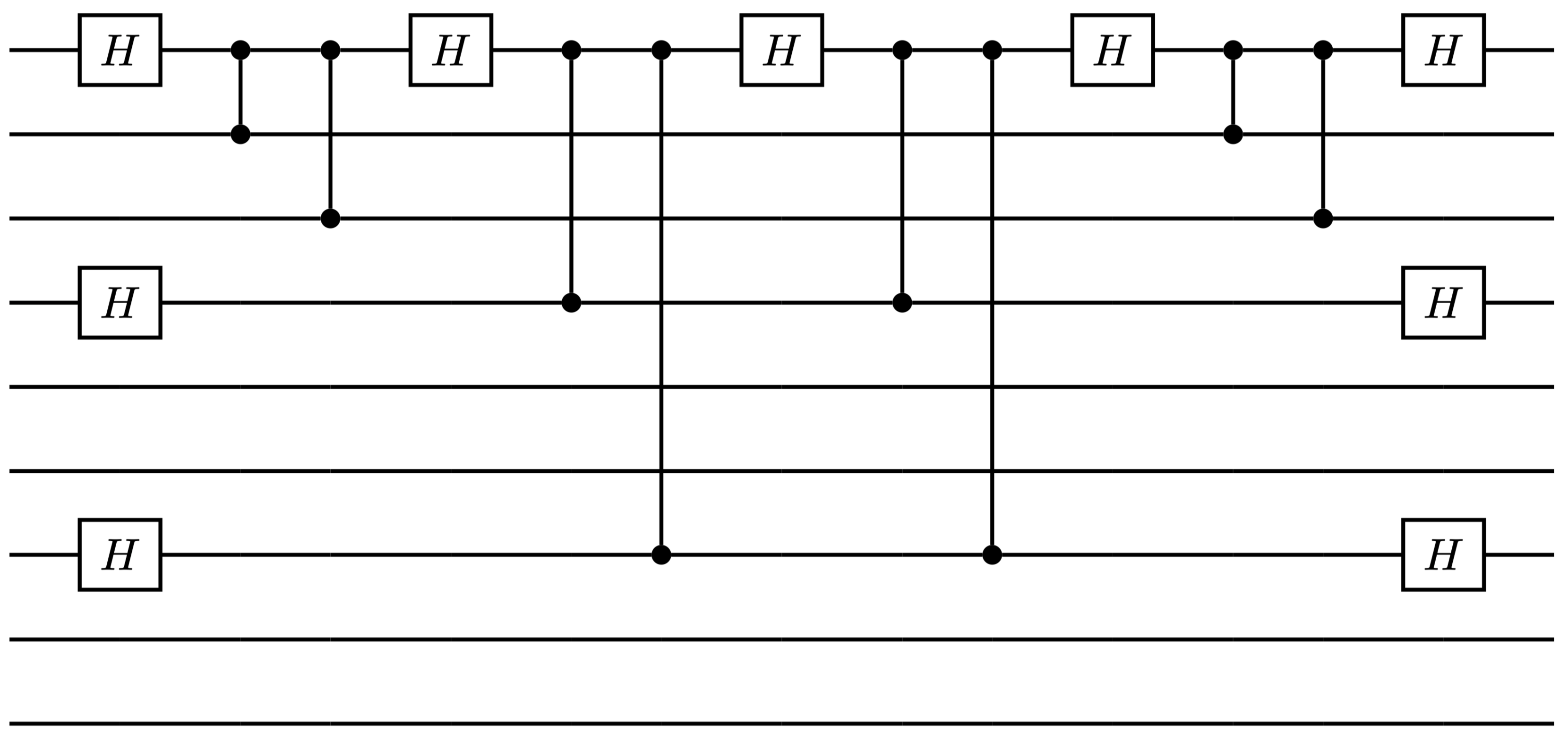}
    \caption{
        \textbf{Circuit to implement the lattice-level Hadamard gate of the (3, 3) parity encoding.}
        The circuit is composed of multiple Hadamard and \cz~gates.
    }
    \label{fig:logical_hadamard_circuit}
\end{figure}

\newpage

\section*{Supplementary Note~8: Algorithm for generating a merging graph of a component}
\label{SN:merging_graph}

We here present a systematic algorithm to construct a merging graph $G_\mr{mrg}$ of a component of a post-$H$ microcluster as follows:
\begin{enumerate}
    \item Initialize the graph $G_\mr{mrg} = (V, E)$ by the graph $G=(V_0, E_0)$ of the component; that is, $V \leftarrow V_0, E \leftarrow E_0$.
    \item Let us define $V_{\mr{deg}\geq3} := \qty{ v \in V \middle| d_v \geq 3 }$. This set is fixed and not updated during the entire process. For each vertex $v \in V_{\mr{deg}\geq3}$, perform the follows:
    \begin{enumerate}
        \item Remove $v$ from $G_\mr{mrg}$ and add $d_v - 1$ new vertices. Let $V_\mr{new} = \qty(v_\mr{new}^{(1)}, \cdots, v_\mr{new}^{(d_v - 1)})$ denote the series of the new vertices and $V_\mr{ngh} = \qty(v_\mr{ngh}^{(1)}, \cdots, v_\mr{ngh}^{(d_v)})$ denote the series of the vertices that were adjacent to $v$ before removing it.
        The order of the vertices in $V_\mr{ngh}$ can be arbitrarily chosen.
        \item Connect the vertices in $V_\mr{new}$ linearly with \textit{internal} edges; namely, connect $\qty(v_\mr{new}^{(1)}, v_\mr{new}^{(2)})$, $\qty(v_\mr{new}^{(2)}, v_\mr{new}^{(3)})$, and so on.
        \item Choose one of the vertices in $V_\mr{new}$ arbitrarily and term it the seed vertex $v_\mr{seed}$ of $v_0$, where $v_0$ is the vertex in $G$ from which $v$ originates.
        \item Let us define a series $V_\mr{new}'$ by omitting $v_\mr{seed}$ from $V_\mr{new}$ while keeping the order of the other vertices. For each $i \in \qty{1, \cdots, d_v-2}$, connect $v_\mr{ngh}^{(i)}$ and the $i$th element of $V_\mr{new}'$ with an \textit{external} edge.
        \item Connect $\qty(v_\mr{new}^{(1)}, v_\mr{ngh}^{(d_v-1)})$ and $\qty(v_\mr{new}^{(d_v-1)}, v_\mr{ngh}^{(d_v)})$ with \textit{external} edges.
    \end{enumerate}
    \item Remove all vertices with degree 1 from the graph. For each vertex $v_0 \in V_0$ with degree 2, if $v$ is the vertex in $V$ originating from $v_0$ (which is not removed in the previous steps), define the seed vertex of $v_0$ as $v$.
\end{enumerate}

It is worth noting that there are two degrees of freedom in the above algorithm for each vertex with a degree larger than 2: (i) the order of the series $V_\mr{ngh}$ and (ii) the selection of the seed vertex.
Different merging graphs can be constructed depending on their selection, which may severely affect the resource overheads.

\newpage

\section*{Supplementary Note~9: Analysis on the optimizing strategy for generating a microcluster}

\begin{figure}[htb!]
    \centering
    \includegraphics[width=\textwidth]{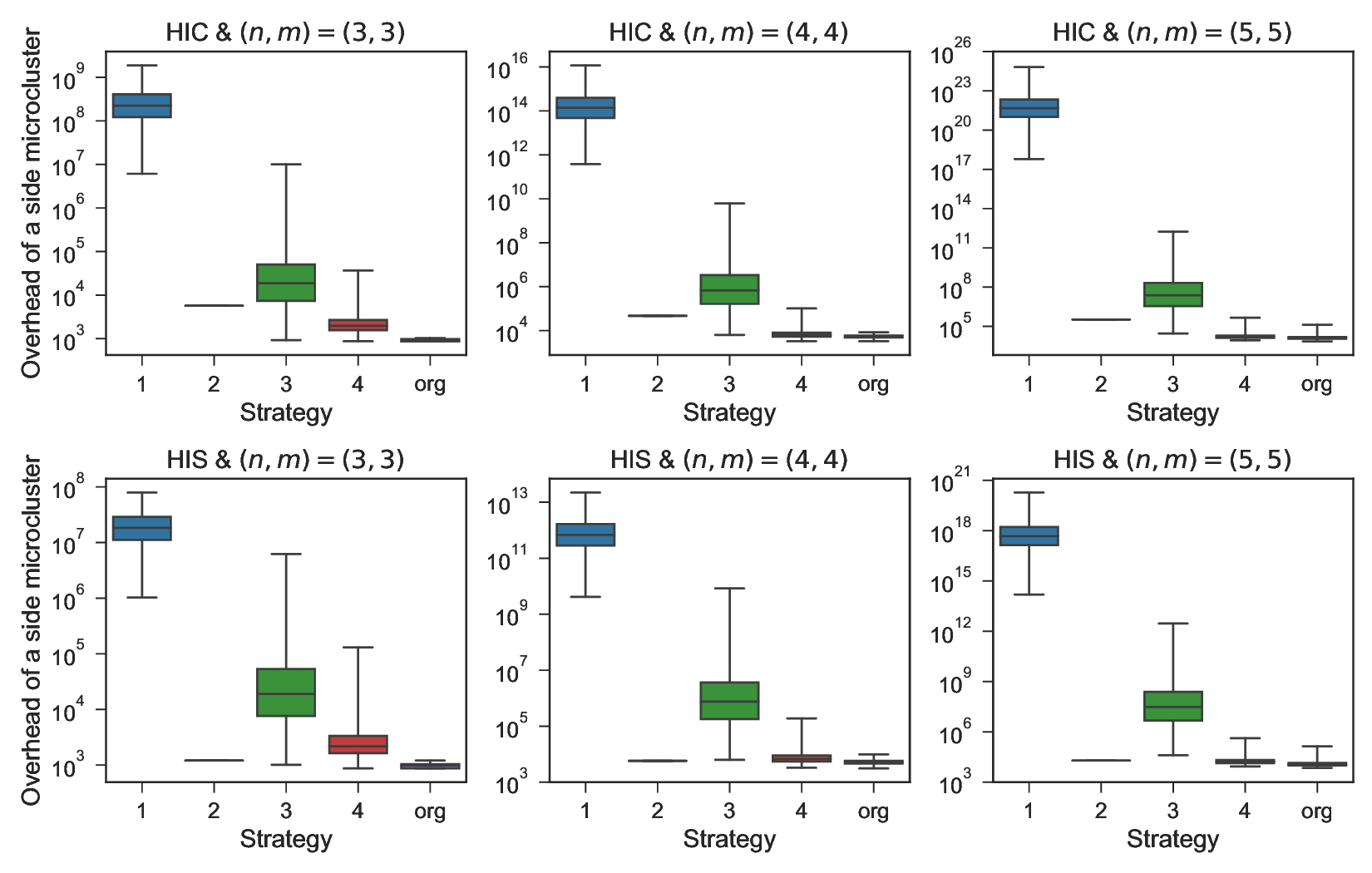}
    \caption{
        \textbf{Comparison of different strategies for the process of generating a post-$H$ microcluster.}
        It shows the distribution of the calculated overhead $N_\mr{GHZ}^\mr{side}$ of a side microcluster depending on the used strategy (among the original strategy and its four variants) for several settings on the $H$-configuration and the values of $n$ and $m$.
        We considered 9,600 samples for each box plot.
        Each box shows the range between the first and third quartile and the line crossing it represents the median.
        The minimum and maximum values are indicated by whiskers.
        Note that, unlike conventional box plots, outliers are not marked and the whiskers visualize the entire data range.
 }
    \label{fig:overhead_strategy_comparison}
\end{figure}

Here, we show evidence that our optimizing strategy for the process of generating a post-$H$ microcluster is indeed highly effective, by comparing its performance with those of its variants constructed by omitting or altering specific steps.
We consider four of such variants that are the same as the original strategy except for the following differences:
\begin{itemize}
    \item (Variant 1) The original physical-level graph is directly used as a single component without decomposition.
    \item (Variant 2) GHZ-$N$ states for $N \geq 3$ are first constructed and then they are merged by fusions to construct a post-$H$ microcluster, which is the process described before introducing merging graphs in the main text. A GHZ-$N$ state is generated by merging two GHZ-$\qty(N/2 + 1)$ states if $N$ is even, or by merging a GHZ-$\qty[(N+1)/2]$ state and a GHZ-$\qty[(N+3)/2]$ state if $N$ is odd. The order of the fusions is determined by the same strategy as the original one, regarding the merging graph where the vertices correspond to general GHZ states and the edges indicate the fusions to perform.
    Accordingly, the weight of each vertex is not initialized to 1, but to the expected number of GHZ-3 states used to generate the corresponding GHZ state.
    \item (Variant 3) The merging order is just randomly chosen without using a specific strategy.
    \item (Variant 4) The merging order is chosen by considering only the weights of the vertices without using the edge coloring algorithm. Namely, steps 2, 3, and 4 in the original strategy in the main text are replaced with contracting a random edge in $E_\mr{min.wgt}$.
\end{itemize}
Supplementary Figure~\ref{fig:overhead_strategy_comparison} displays how the overhead of a side microcluster (namely, the expected number $N_\mr{GHZ}^\mr{side}$ of GHZ-3 states required to generate it) varies depending on the used strategy for several settings.
Since the strategies contain randomness, the distributions of the outcomes are visualized as box plots.
It is clearly shown that the original strategy is the most optimal in general, although Variant~2 or~4 is also as effective as the original strategy for some cases. 
Moreover, the original strategy gives the least variance on the calculated overhead (except for Variant~2), which means that the optimal point can be found quickly.

\newpage

\section*{Supplementary Figure~8}

\begin{figure}[htb!]
    \centering
    \includegraphics[width=\textwidth]{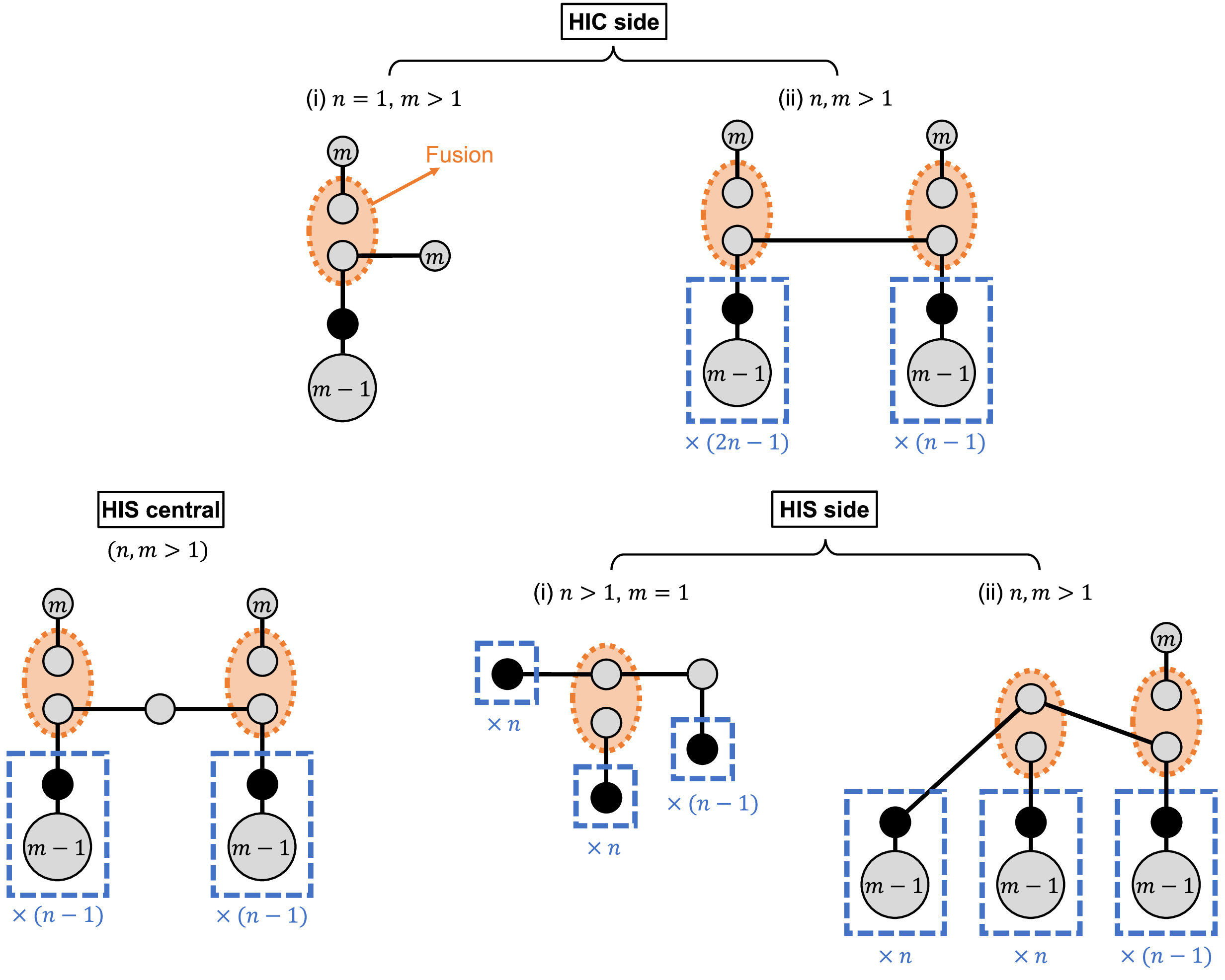}
    \caption{
        \textbf{Decomposition of various post-$H$ microclusters.}
        Different types of post-$H$ microclusters are decomposed by the method shown in Fig.~9 of the main text.
        The types of post-$H$ microclusters that are not presented here (including the central microcluster of HIC and the other microclusters that do not meet the presented conditions) do not have connected pairs of recurrent subgraphs, thus their physical-level graphs are single components by themselves.
    }
    \label{fig:microcluster_decompositions}
\end{figure}

\newpage

\section*{Supplementary Table~1}

\begin{table}[htb!]
    \caption{
        \textbf{Information of the data points along the upper envelope lines in Fig.~6 of the main text.}
        $N_\mr{GHZ}^*$, $\mathcal{N}_{10^{-7}}$, and $d_{10^{-7}}$ at $\eta = 1\%$ are not calculated when $\eta_\mr{th} < 1\%$.
    }
    \label{table:optimal_paramter_settings}
    \centering
    \begin{ruledtabular}
    \begin{tabular}{ccccccccc|ccccccccc}
        $\eta_\mr{th}$ & \begin{tabular}[c]{@{}c@{}} $N_\mr{GHZ}^*$ \\ ($\eta = \eta_\mr{th}/2$) \end{tabular} & \begin{tabular}[c]{@{}c@{}} $N_\mr{GHZ}^*$ \\ ($\eta = 1\%$) \end{tabular} & \begin{tabular}[c]{@{}c@{}} $\mathcal{N}_{10^{-7}}$ \\ ($\eta = 1\%$) \end{tabular} & $d_{10^{-7}}$ & $n$ & $m$ & $j$ & $H$-config. & $\eta_\mr{th}$ & \begin{tabular}[c]{@{}c@{}} $N_\mr{GHZ}^*$ \\ ($\eta = \eta_\mr{th}/2$) \end{tabular} & \begin{tabular}[c]{@{}c@{}} $N_\mr{GHZ}^*$ \\ ($\eta = 1\%$) \end{tabular} & \begin{tabular}[c]{@{}c@{}} $\mathcal{N}_{10^{-7}}$ \\ ($\eta = 1\%$) \end{tabular} & $d_{10^{-7}}$ & $n$ & $m$ & $j$ & $H$-config. \\ \hline
\multicolumn{9}{c|}{\textbf{Single-photon resolving detector with post-selection}} &\multicolumn{9}{c}{\textbf{Single-photon resolving detector without post-selection}} \\
0.009 & $3.3 \times 10^2$ & $3.7 \times 10^2$ &  &  & 1 & 4 & 3 & HIC & 0.009 & $7.2 \times 10^2$ &  &  &  & 3 & 2 & 1 & HIC \\
0.02 & $3.9 \times 10^2$ & $3.9 \times 10^2$ & $2.1 \times 10^{7}$ & 21 & 2 & 2 & 1 & HIC & 0.015 & $8.4 \times 10^2$ & $8.6 \times 10^2$ & $2.5 \times 10^{8}$ & 37 & 2 & 3 & 2 & HIC \\
0.03 & $8.8 \times 10^2$ & $8.2 \times 10^2$ & $1.5 \times 10^{6}$ & 7 & 3 & 2 & 1 & HIC & 0.022 & $1.6 \times 10^3$ & $1.6 \times 10^3$ & $6.8 \times 10^{6}$ & 9 & 2 & 4 & 3 & HIC \\
0.035 & $1.0 \times 10^3$ & $9.2 \times 10^2$ & $3.8 \times 10^{6}$ & 9 & 2 & 3 & 2 & HIC & 0.023 & $2.3 \times 10^3$ & $2.3 \times 10^3$ & $1.4 \times 10^{10}$ & 101 & 5 & 2 & 1 & HIC \\
0.036 & $1.8 \times 10^3$ & $1.6 \times 10^3$ & $3.0 \times 10^{6}$ & 7 & 4 & 2 & 1 & HIC & 0.024 & $2.6 \times 10^3$ & $2.6 \times 10^3$ & $4.9 \times 10^{6}$ & 7 & 2 & 5 & 4 & HIC \\
0.04 & $1.9 \times 10^3$ & $1.7 \times 10^3$ & $3.2 \times 10^{6}$ & 7 & 2 & 4 & 3 & HIC & 0.043 & $4.4 \times 10^3$ & $3.9 \times 10^3$ & $2.6 \times 10^{6}$ & 5 & 4 & 3 & 1 & HIC \\
0.052 & $2.7 \times 10^3$ & $2.1 \times 10^3$ & $1.4 \times 10^{6}$ & 5 & 3 & 3 & 1 & HIC & 0.048 & $7.1 \times 10^3$ & $6.0 \times 10^3$ & $4.0 \times 10^{6}$ & 5 & 5 & 3 & 1 & HIC \\
0.067 & $5.3 \times 10^3$ & $3.9 \times 10^3$ & $5.2 \times 10^{5}$ & 3 & 4 & 3 & 1 & HIC & 0.05 & $8.7 \times 10^3$ & $7.3 \times 10^3$ & $4.9 \times 10^{6}$ & 5 & 5 & 3 & 1 & HIS \\
0.074 & $8.6 \times 10^3$ & $6.1 \times 10^3$ & $8.1 \times 10^{5}$ & 3 & 5 & 3 & 1 & HIC & 0.052 & $9.3 \times 10^3$ & $9.8 \times 10^3$ & $1.3 \times 10^{6}$ & 3 & 4 & 4 & 2 & HIC \\
0.085 & $2.3 \times 10^4$ & $1.5 \times 10^4$ & $2.0 \times 10^{6}$ & 3 & 5 & 4 & 2 & HIC & 0.054 & $1.4 \times 10^4$ & $1.1 \times 10^4$ & $1.5 \times 10^{6}$ & 3 & 4 & 5 & 3 & HIC \\
&&&&&&&& & 0.061 & $1.9 \times 10^4$ & $1.4 \times 10^4$ & $1.9 \times 10^{6}$ & 3 & 5 & 4 & 2 & HIC \\
&&&&&&&& & 0.063 & $2.3 \times 10^4$ & $1.7 \times 10^4$ & $2.3 \times 10^{6}$ & 3 & 5 & 5 & 3 & HIC \\
\hline
\multicolumn{9}{c|}{\textbf{On-off detector with post-selection}} &\multicolumn{9}{c}{\textbf{On-off detector without post-selection}} \\
0.009 & $1.8 \times 10^3$ &  &  &  & 2 & 3 & 2 & HIC & 0.005 & $1.9 \times 10^3$ &  &  &  & 3 & 3 & 2 & HIC \\
0.012 & $3.6 \times 10^3$ & $3.9 \times 10^3$ & $2.1 \times 10^{9}$ & 45 & 2 & 4 & 2 & HIC & 0.008 & $2.3 \times 10^3$ &  &  &  & 3 & 3 & 1 & HIS \\
0.013 & $4.6 \times 10^3$ & $5.0 \times 10^3$ & $4.6 \times 10^{8}$ & 25 & 2 & 5 & 4 & HIC & 0.013 & $3.7 \times 10^3$ & $3.9 \times 10^3$ & $2.1 \times 10^{8}$ & 21 & 4 & 3 & 1 & HIC \\
0.022 & $1.0 \times 10^4$ & $1.0 \times 10^4$ & $1.1 \times 10^{9}$ & 27 & 3 & 3 & 1 & HIC & 0.014 & $4.4 \times 10^3$ & $4.6 \times 10^3$ & $6.6 \times 10^{8}$ & 29 & 3 & 4 & 2 & HIS \\
0.024 & $1.1 \times 10^4$ & $1.0 \times 10^4$ & $2.0 \times 10^{7}$ & 7 & 3 & 4 & 2 & HIC & 0.016 & $4.7 \times 10^3$ & $4.8 \times 10^3$ & $1.6 \times 10^{9}$ & 39 & 4 & 3 & 1 & HIS \\
0.035 & $3.1 \times 10^4$ & $2.6 \times 10^4$ & $1.7 \times 10^{7}$ & 5 & 4 & 4 & 2 & HIS & 0.02 & $6.0 \times 10^3$ & $6.0 \times 10^3$ & $4.6 \times 10^{7}$ & 11 & 5 & 3 & 1 & HIC \\
0.044 & $2.4 \times 10^5$ & $1.9 \times 10^5$ & $1.2 \times 10^{8}$ & 5 & 5 & 4 & 1 & HIC & 0.023 & $9.6 \times 10^3$ & $9.3 \times 10^3$ & $7.1 \times 10^{7}$ & 11 & 4 & 4 & 2 & HIS \\
&&&&&&&& & 0.025 & $1.4 \times 10^4$ & $1.4 \times 10^4$ & $5.8 \times 10^{7}$ & 9 & 4 & 5 & 3 & HIS \\
&&&&&&&& & 0.028 & $1.5 \times 10^4$ & $1.5 \times 10^4$ & $6.3 \times 10^{7}$ & 9 & 5 & 4 & 2 & HIC \\
&&&&&&&& & 0.03 & $1.6 \times 10^4$ & $1.7 \times 10^4$ & $1.2 \times 10^{8}$ & 11 & 5 & 4 & 1 & HIS \\
&&&&&&&& & 0.032 & $2.3 \times 10^4$ & $2.1 \times 10^4$ & $8.8 \times 10^{7}$ & 9 & 5 & 5 & 2 & HIS \\
    \end{tabular}
    \end{ruledtabular}
\end{table}

\bibliography{references}